\documentclass[useAMS, usenatbib]{mn2e}
\usepackage{amsmath}
\usepackage{xspace}
\usepackage{psfig}
\usepackage{psfig}
\usepackage{defs}

\def\eq{equation}
\def\fig{Fig.}
\def\figs{Figs.}

\def\etal{{\it et al.~\/}}
\def\cf{{\it cf.}}
\def\ie{{\it i.e.}}
\def\eg{{\it e.g.}}
\def\ltsima{$\; \buildrel < \over \sim \;$}
\def\simlt{\lower.5ex\hbox{\ltsima}}
\def\gtsima{$\; \buildrel > \over \sim \;$}
\def\simgt{\lower.5ex\hbox{\gtsima}}

\def\h2{H$_2$\xspace}

\def\Mpc{h$^{-1}$ Mpc\xspace}
\def\kpc{h$^{-1}$ kpc\xspace}
\title{Dependence of the Inner DM Profile on the Halo Mass}

\author[M. Ricotti]
{Massimo Ricotti\\ 
Institute of Astronomy, Madingley Road, Cambridge CB3 0HA\\
ricotti@ast.cam.ac.uk}
\date{Accepted ---. Received ---; in original form 10 December 2002}
\pagerange{\pageref{firstpage}--\pageref{lastpage}}
\pubyear{2002}
\begin{document}
\maketitle
\label{firstpage}

\begin{abstract}
  I compare the density profile of dark matter (DM) halos in cold dark
  matter (CDM) N-body simulations with 1 \Mpc, 32 \Mpc, 256 \Mpc and
  1024 \Mpc box sizes. I compare the profiles when the most massive
  halos are composed of about $10^5$ DM particles. The DM density
  profiles of the halos in the 1 \Mpc box at redshift $z \sim 10$ show
  systematically shallower cores with respect to the corresponding
  halos in the 32 \Mpc simulation at $z \sim 3$ that have masses,
  $M_{dm}$, typical of the Milky Way and are fitted by a NFW profile.
  The DM density profiles of the halos in the 256 \Mpc box at $z \sim
  0$ are consistent with having steeper cores than the corresponding
  halos in the 32 \Mpc simulation, but higher mass resolution
  simulations are needed to strengthen this result.  Combined, these
  results suggest that the density profile of DM halos is not
  universal, presenting shallower cores in dwarf galaxies and steeper
  cores in clusters. More work is needed to validate this finding
  at $z=0$.  Physically the result sustains the hypothesis that the
  mass function of the accreting satellites determines the inner slope
  of the DM profile. But the result can also be interpreted as a trend
  with the dynamical state in the assembly process of halos of different
  mass.  In comoving coordinates, $r$, the profile
\[
\rho_{dm} \propto {1 \over X^\alpha(1+X)^{3-\alpha}}
\]
with $X=c_\Delta r/r_\Delta$ and $\alpha =(9+3n)/(5+n) \approx
1.3+(M_{dm,14}^{1/6}-1)/(M_{dm,14}^{1/6}+1)$, provides a good fit to
all the DM halos from dwarf galaxies to clusters at any redshift with
the same concentration parameter $c_\Delta \sim 7$. Here,
$r_\Delta(M_{dm})$ is the virial radius, $n$ is the effective spectral
index of the initial power spectrum of density perturbations and
$M_{dm,14}=M_{dm}/(3 \times 10^{14}$ M$_\odot$). The slope, $\gamma$,
of the outer parts of the halo appears to depend on the acceleration
of the universe: when the scale parameter is $a=(1+z)^{-1} \simlt 1$,
the slope is $\gamma \approx 3$ as in the NFW profile, but $\gamma
\approx 4$ at $a >1$ when $\Omega_\Lambda \sim 1$ and the universe is
inflating. The shape of the DM profiles presents a significant scatter
around the mean. It is therefore important to analyse a significant
statistical sample of halos in order to determine the mean profile.
  
I compare the DM profiles in the 1 \Mpc box with the same
simulation including stars, baryons and radiative transfer presented
by Ricotti, Gnedin and Shull. Radiative feedback effects produce a
larger scatter in the density profile shapes but, on average, do not
affect the shape of the DM profiles significantly.
\end{abstract}

\section{Introduction}

According to the currently favoured galaxy formation scenarios,
galaxies are formed from the gravitational growth of tiny density
perturbations imprinted on a uniform universe during inflation. Small
mass perturbations grow faster and constitute the building blocks for
the assembly of larger galaxies and clusters. In the nonlinear phase
of the gravitational collapse the dark matter (DM) particles are shock
heated to the virial temperature and settle into a dark halo with mean
overdensity about $\Delta=178$ (according to the simple spherical
collapse model in a flat $\Omega_0=1$ universe).  Analysing N-body
simulations of hierarchical structure formation
\cite*{Navarro:96,Navarro:97} (hereafter, NFW) have proposed that DM
halos develop a universal density profile, valid for virialized halos
at any redshift and with any mass, from dwarf galaxies to clusters.
Further works have shown that an approximatively universal profile
develops regardless of the details of the adopted cosmology and
initial power spectrum of density perturbations
\citep{Huss:99,Eke:01}. The NFW density profile has a core cusp $\rho
\propto r^{-1}$ and in the outer regions the density decreases as
$\rho \propto r^{-3}$. The details of the gravitational collapse
affect only the relative size of the core with respect to the virial
radius (the halo concentration parameter, $c_\Delta$) but not the
slope of the profile. This result implies that during the
virialization process the memory of the mass function and profile
shapes of the building blocks that formed the halo is lost.

Recent observations of dwarf galaxies (spheroidal and irregular) and
low surface brightness (LSB) galaxies seem to show that a flat core
for the DM density profile is more compatible with the measurements of
the rotation curves \citep{vandenBosch:00,deBlok:01,deBlok:02,
  BorrielloS:01,SalucciB:00}.  These observations have renewed
interest in the subject and a number of solutions have been proposed
to solve this possible discrepancy.  The proposed solutions range from
feedback effects of the baryons and stars on the DM density profile to
modifications of the physical properties (\eg, scattering cross
section, temperature) of the DM particles. Some examples of
alternatives to CDM are warm dark matter \citep{BodeO:01} and
self-interacting DM \citep{Spergel:00}. The potential discrepancy
between the steep DM cores found in N-body simulations with the
observed flat core of dwarf galaxies is perhaps the most serious
problem faced in CDM cosmologies today.

High resolution simulations and careful convergence tests have been
done or are underway to understand if the discrepancies found by some
authors (\eg, \cite{Moore:98,Moore:99} find steeper core profiles) are
due to numerical artifacts.  In the core of DM halos the particle
crossing time is much shorter than the Hubble time and the integration
of trajectories is subject to subtle numerical effects that are
difficult to keep under control.  \cite{Power:02} estimated that about
1 million particles are needed in order to have convergent results in
the inner 1\% of the halo virial radius.  To have such a high number
of particles per halo the use of a special technique is required. This
technique consists of simulating with high mass resolution only the
particles that will end up in the halo of interest at $z=0$, while
having less mass resolution for the particles that end up far away
from the halo of interest. This has the drawback that each simulation
can resolve only one halo at a time, therefore selection effect biases
and a poor statistical sample could affect the reliability of the
final result even if the simulation is very accurate and has high
resolution.  Note that most of the detailed and computationally
expensive work on the subject has focused on simulating halos at $z=0$
with typical mass similar to the Milky Way.

The mass function of the building blocks of dwarf galaxies is
different in comparison to the one for larger galaxies or clusters. On
small scales (or masses) the mass fraction of virialized halos is
about constant as a function of the logarithmic DM mass of the halos
because the power spectrum has a slope $n \sim -3$.  This means that
small and large mass satellites contribute equally to the mass of the
accreting halo.  In massive halos as in the Milky Way or in clusters
of galaxies, the contribution from large mass accreting satellites is
instead dominant.  For this reason it is more crucial to resolve very
small mass satellites (\ie, to have high mass resolution) in
simulating dwarf galaxies than in simulating massive galaxies or
clusters.

In this paper I compare the density profiles of DM halos in four CDM
N-body simulations that are identical apart from the box sizes that
are 1 \Mpc, 32 \Mpc, 256 \Mpc and 1024 \Mpc.  The simulation method is
a two-level particle-particle particle-mesh (P$^3$M) with $256^3$
particles and $\Lambda$CDM cosmology.  In dimensionless units the only
difference between the simulations is the initial power spectrum of
density perturbations. I compare the profiles when the most massive
halos are composed of about $10^5$ DM particles and the
clustering in the simulations is similar (see \fig~\ref{fig:hal}).
This is done in order to minimise possible systematic errors in the
simulations and to be more confident that any variation in the profile
shapes is produced by the different initial power spectrum.  The
masses of the halos studied in the four simulations, from the small to
the large box, are typical of dwarf galaxies, Milky Way mass galaxies,
clusters and superclusters of galaxies, respectively.

\begin{figure}
\centerline{\psfig{figure=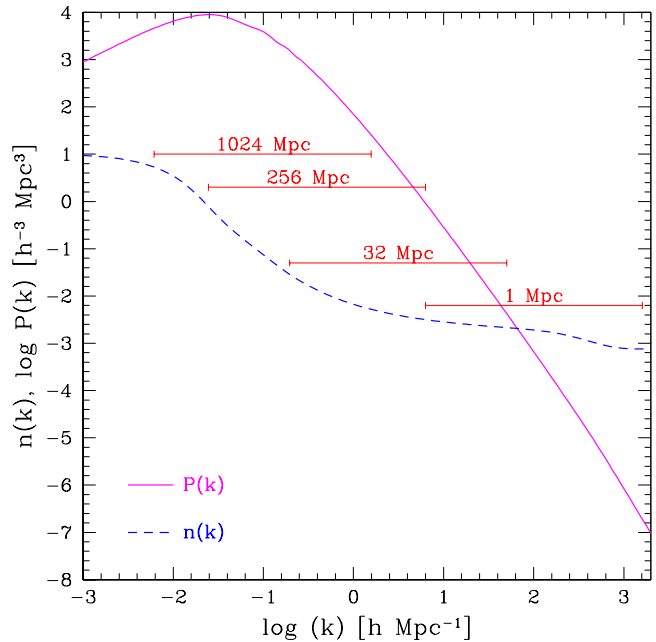,width=9cm}}
\caption{\label{fig:pk} Linear power spectrum of density
  fluctuations at $z=0$ (solid line) and logarithmic slope, $n(k)$, of the
  power spectrum (dashed line). The horizontal bars show the range
  of wave numbers, $k=2 \pi/\lambda$, resolved in the $1, 32, 256$ and $1024$
  \Mpc simulations.}
\end{figure}

If the cores of small mass halos are dense enough to resist disruption
and survive ``undigested'' when incorporated into a larger object,
they should determine the halo structure in the inner regions. Using
scaling arguments \cite*{SubramanianCO:00} have shown that, in a flat
universe with an initial power spectrum of density perturbation $P(k)
\propto k^n$, the core density profile is $\rho_{dm} \propto r^\alpha$
with $\alpha = (9+3n)/(5+n)$. The numerical results that they report
approximately confirm this expectation. In $\Lambda$CDM cosmologies
the logarithmic slope of the initial power spectrum ranges between
$-3<n<1$, depending on the mass scale (see \fig~\ref{fig:pk}). From
dwarfs to superclusters the logarithmic slope of the core density
profile should range between $0 \simlt \alpha \simlt 2$. As shown in
\fig~\ref{fig:pk}, the power spectrum in the 1 \Mpc box is a power law
with $n \approx -2.7$.  According to the aforementioned scaling
relationship $\alpha \approx 0.4$. This slope is in agreement with
what I find in the 1 \Mpc box.  In the 32 \Mpc box the power spectrum
is not a single power law but the slope is close to $n \approx -2 $
and therefore $\alpha \approx 1$ is expected, consistent with the NFW
universal profile and with the result found for this simulation. In
cluster and supercluster size halos $\alpha$ should be larger than in
galaxy size halos ($\alpha \sim 1.5$).  Indeed, in the 256 \Mpc
simulation I find slopes of the inner profile $\alpha \sim 1.4$. But
this is not a strong result because the number of particles in each
halo is barely sufficient to start seeing a deviation with respect to
the NFW profile.  Moreover, the introduction of the parameter $\alpha$
for the slope of the inner part of the halo profile (which can be
easily calculated knowing the mass of the galaxy halo) does not
increase the number of adjustable parameters in the profile fitting
formula. This happens because the concentration parameter,
$c_\Delta^{NFW}(M_{dm})$, needed to fit the profiles with the NFW
formula, is in this case a constant.  The changing slope of the inner
profile mimics the variation of the concentration parameter in the NFW
profile, shifting the value of the radius where the circular velocity
reaches its maximum value (see appendix~\ref{app}).

\begin{table*}
\centering
\caption{Clustering and time steps in the four simulations. $a$ is the
  scale factor; $N_p$ has two columns: the numbers on the left show
  the number of 
  particles in the most massive halo, the right contains the mean number
  of particles for the ten most massive halos.}
\label{tab:one}
\begin{tabular*}{14cm}[]{l|cc||l|cc||l|cc||l|cc}
\multicolumn{3}{c}{1 \Mpc} \vline \vline &
\multicolumn{3}{c}{32 \Mpc} \vline \vline &
\multicolumn{3}{c}{256 \Mpc} \vline \vline &
\multicolumn{3}{c}{1024 \Mpc} \\
\hline
a & steps & $N_p/10^4$ & a & steps & $N_p/10^4$ & a &steps & $N_p/10^4$ & a &  steps & $N_p/10^4$ \\
\hline
0.07 & 1213 & 5.0 2.5 & 0.200 & 1530 & ..  ..  & 0.9 & 2419 & 4.0 3.0 & 2.3
& 1530 & 0.16 0.14\\
0.08 & 1443 & 5.5 3.3 & 0.230 & 1814 & 5.5 4.0 & 1.0 & 2546 & 4.0 3.0 & 4.0 
& 1681 & 0.17 0.15 \\
0.085& 1542 & 10 5.0  & 0.250 & 1974 & 4.5 4.0 & 1.2 & 2800 & 5.0 4.0 & 10
& 1809 & 0.17 0.15\\
0.090& 1678 & 8.3 5.3 & 0.270 & 2122 & 13 6.0  & 1.5 & 3078 & 7.5 5.0 & 100
& 1880 & 0.17 0.15\\
0.095& 1804 & 13 6.5  & 0.300 & 2470 & 18 7.5  & 1.7 & 3196 & 7.5 5.5 &1000
& 1910 & .. .. \\
\end{tabular*}
\end{table*}
This paper is organised as follows: in \S~\ref{sec:sim} I explain the
methods for the simulations and in \S~\ref{sec:res} I show the results
for dwarf galaxies, normal galaxies and clusters of galaxies. In
\S~\ref{sec:prof} I demonstrate how a density profile with changing
inner slope $\alpha$ can fit all the profiles from dwarfs to clusters
with a constant concentration parameter. The discussion and
conclusions are presented in \S~\ref{sec:con}.  The equations for the
circular velocity and integrated mass for the NFW profile and for a
halo with arbitrary inner and outer slopes of the density profile are
contained in appendix~\ref{app}.

\section{The Method}\label{sec:sim}

In the literature most efforts have focused on studying galaxies at
$z=0$ with masses typical of the Milky Way.
\cite{RicottiGSa:02,RicottiGSb:02} have performed high-resolution
simulations of the formation of the first galaxies using a
cosmological code that solves the equations for the DM particles,
baryons, stars and radiative transfer. These simulations achieve a
mass resolution of $M_p =4.9 \times 10^3$ M$_\odot$ for the DM using
$256^3$ particles and 1 \Mpc box size. In these simulations at $z \sim
10$ or, expressing the redshift in terms of the scale factor, at
$a=(1+z)^{-1} \sim 0.09$, the profiles of the most massive halos
($M_{dm} \sim 2 \times 10^8$ M$_\odot$) appear flatter than the NFW
profile.  I therefore perform the same simulation for only the DM
particles, finding again systematically flatter halo cores than found
by NFW. Since this result could be produced by subtle numerical
problems of the N-body integration or the small number of particles in
the halo (typically between $10^4$ and $10^5$), I repeat the
simulation for a 32 \Mpc box down to $a=0.3$ when the most massive
halos have the same number of DM particles as in the 1 \Mpc box at
$a=0.09$.  In this case the profiles are well fitted by the NFW
universal profile. The differences of the profiles in this two
simulations are likely to be caused by the different initial power
spectrum of density perturbations, since the they have identical
smoothing length in dimensionless units, identical Fourier modes of
the initial density perturbations, similar number of integration time
steps and comparable number of particles per halo. According to this
paradigm and the results of \cite{SubramanianCO:00}, the slope of the
DM profiles in simulations with box sizes larger than 32 \Mpc should
be steeper than the NFW profiles.  I perform two additional
simulations with 256 \Mpc and 1024 \Mpc box sizes to test this
hypothesis. The large box simulations present the problem that in
order to achieve the same degree of clustering as in the smaller boxes
it is necessary to evolve the simulation in the future when $a \simgt
1$ and $\Omega_\Lambda > 0.7$.  For the 256 \Mpc simulation the
required level of clustering is reached when $0.9 < a <1.5$ and the
number of particles per halo is barely sufficient to start noticing a
steepening of the inner profile with respect to the NFW profile. In
the case of the 1024 \Mpc box the level of clustering reached in the
smaller boxes can never be achieved because the universe begins to
inflate (the scale factor becomes $a>1000$ in a few time steps) and
the event horizon starts decreasing, preventing any further
clustering. I present the results for this simulation because they are
interesting for understanding what determines the outer slope of the
density profile.  The number of particles in these halos is too small
for measuring the slope of the inner profile but it is evident that,
in an accelerating universe, the outer slope of the density profile is
steeper. Analytically, \cite{SubramanianCO:00} show that the slope of
the outer profile is $\gamma \sim 4$ in a low density universe and in
an accelerating $\Lambda$CDM universe. Table~\ref{tab:one} lists the
scale factors when the simulations with different box sizes,
$L_{box}$, have run for a comparable number of time steps and the most
massive halos are composed of the same number of DM particles.
 
In the following paragraphs I explain in greater detail the numerical
techniques and the sanity checks adopted to derive the density
profiles and the rotation curves.

\subsection{N-body Simulations}

I use an N-body code based on the P$^3$M method with two levels of
mesh refinement \citep{Bertschinger:91,GnedinB:96}. In all the
simulations the number of particles is $256^3$ and the Plummer
softening parameter is $\epsilon=0.02$ in cell units $\Delta L$, where
$\Delta L=L_{box}/256$ and $L_{box}$ is the box size. Unless stated
otherwise, I use scale-free (dimensionless) units in this paper.  The
cosmology adopted is a flat $\Lambda$CDM model with
$\Omega_\Lambda=0.7$, $\Omega_m=0.3$ and $h=0.7$. The initial
conditions for the density and velocity fields are calculated using the
COSMIC package \citep{Bertschinger:95} assuming a scale-invariant
(tilt $n=1$) initial spectrum of density perturbations and
$\Lambda$CDM transfer function.  The spectrum, shown in
\fig~\ref{fig:pk}, is normalised imposing a variance $\sigma_8=0.91$
in spheres with radius of 8 \Mpc at $z=0$. The initial conditions are
calculated using the Zel'dovich approximation until $z=166, 91, 48$ and
$26$ for the $L_{box}=1, 32, 256$ and $1024$ \Mpc boxes, respectively.
The N-body simulations start at those redshifts. I use the same random
realization for the wave phases and directions in all simulations.

The halos are identified using DENMAX \citep{Bertschinger:91} with
smoothing parameter $G=1000$.  The DENMAX algorithm identifies halos
as maxima of the smoothed density field with smoothing length $L_G =
L_{box}/G$. The algorithm assigns each particle in the simulation to a
group (halo) by moving each particle along the gradient of the density
field until it reaches a local maximum. Unbound particles are then
removed from the group.  The results of DENMAX depend on the degree of
smoothing used to define the density field. A finer resolution in the
density field will split large groups into smaller subunits and vice
versa. This arbitrariness in results is a common problem of any
group-finding algorithm, and it is not only a numerical problem but
often a real physical ambiguity.  Especially at high redshift, since
the merger rate is high, it is difficult to identify or define a
single galaxy halo.  In appendix~B in \cite{RicottiGSa:02} the mass
functions obtained using DENMAX with different smoothing lengths are
compared with the analytical expectation using the Press-Schechter
formalism. It was found that the smoothing parameter $G=1000$ gives
the best results for boxes with $256^3$ particles.  

\begin{figure}
\centerline{\psfig{figure=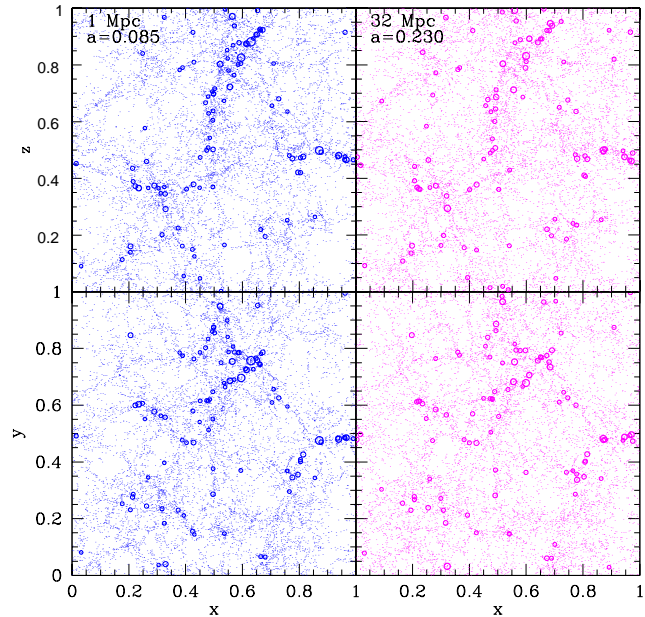,width=9cm}}
\caption{\label{fig:hal} Position of DM halos in the 1 \Mpc box at $a=0.085$
  (left panels) and 32 \Mpc box at $a=0.23$ (right panels). The
  circles show the sizes (virial radii) and positions of the 100 most
  massive halos in the simulation. The 10 most massive halos are
  shown as thicker circles. The positions of the most massive
  10000 halos are shown as points.}
\end{figure}

In each simulation with a different box size the profiles of the halos
are compared when the level of clustering is similar and the positions
of the most massive halos coincide. This happens after a comparable
number of time steps, but the number of steps required becomes larger
as the box size increases since the halos are more concentrated. The
most massive halos in the simulations are composed of about $N_p=10^4
- 10^5$ DM particles and the typical number of time steps to reach
this level of clustering is a few thousands (see Table~\ref{tab:one}).
\fig~\ref{fig:hal} shows the sizes and positions of bound halos in
the 1 \Mpc box at $a=0.085$ and in the 32 \Mpc box at $a=0.230$, when
the clustering in the two simulations is similar.

\subsection{Density Profiles}

The density profiles are calculated in three different ways:\\
(i) Using the list of particles identified as halo members by DENMAX. This means all the bound particles.\\
(ii) Using, additionally, the particles identified by DENMAX as unbound
particles.\\
(iii) Using all the particles. This includes all the halo satellites
and neighbouring halos.\\
Usually profiles using (i) and (ii) are almost indistinguishable. But the
profiles using (iii) look different in the outer parts if the halo is
disturbed by accreting satellites. The profiles obtained with method
(iii) are used for the analysis as in previous studies. The other
profiles are useful to discriminate between isolated halos and halos
that are undergoing accretion of massive satellites. One can choose
whether to exclude or include in the analysis halos that are
undergoing major mergers; either choice does not change the result
significantly.

The profiles are shown in dimensionless units: the radii are expressed
in number of cells $x$ (the cell size is $\Delta L =L_{box}/256$ in
comoving coordinates) and the masses in number of DM particles $N_p$.
The mass and density of each halo are sampled as a function of the
radius, $x$, in shells with constant logarithmic spacing $dx/x=0.01$.
The dimensionless circular velocity is more appropriate to compare the
profiles since the virial radii are difficult to determine when the
halos are not isolated. Moreover, the relative error on the density
profile is smaller at radii corresponding to the maximum of the
circular velocity since the number of particles per shell at
intermediate radii is maximum.
Therefore, if the halo profile is not disturbed by accreting
satellites, the location and the value of the maximum circular
velocity provides the best method to compare halos of different
masses.  The dimensionless circular velocity is defined as,
\[
v_c = {V_c(x) \over V_c^0} = \sqrt{N_{dm}(x) \over x}
\]
where $V_c^0= (G M_p/\Delta L)^{1/2}=7.365 \times 10^{-2}
(L_{box}/1~h^{-1}{\rm Mpc})$ km s$^{-1}$.  If the slope of the inner
density profile is $\rho(r) \propto r^{-\phi}$, the circular velocity
has a slope $V_c \propto r^{(1-\phi/2)}$.  The Poisson error for the
density is $\delta \rho /\rho =dN_p(x)^{-1/2}$, where $dN_p(x)$ is the
number of particles in each shell of radius $x$. The error on the mass
as a function of the radius is $\delta M_{dm} /M_{dm}(x)=N_p(x)^{-1/2}$,
where $N_p(x)$ is the number of particles in each sphere of radius
$x$. The error on the circular velocity is $\delta v_c / v_c= 0.5
\delta M_{dm} /M_{dm}$.  Table~\ref{tab:two} lists the values of
useful constants for converting dimensionless units into physical
units.

\begin{table}
\centering
\caption{Scaling constants.}\label{tab:two}
\begin{tabular*}{8.0 cm}[bt]{c|c|c|c|c}
L$_{box}$ & $\Delta L$ & Log M$_{p}$ & $V_c^0/100$ & a\\
\Mpc & \kpc & h$^{-1}$ M$_\odot$ & km s$^{-1}$ & range\\
\hline
1   & 3.9  & $3.69$ & 7.26  & 0.07-0.09 \\
32  & 125  & $8.21$ & 235.7 & 0.23-0.3 \\
256 & 1000 & $10.9$ & 1885 & 0.9-1.5 \\
1024& 4000 & $12.7$ & 7542 & $> 1000$ \\
\end{tabular*}
\end{table}

\section{Results}\label{sec:res}

\begin{figure}
\centerline{\psfig{figure=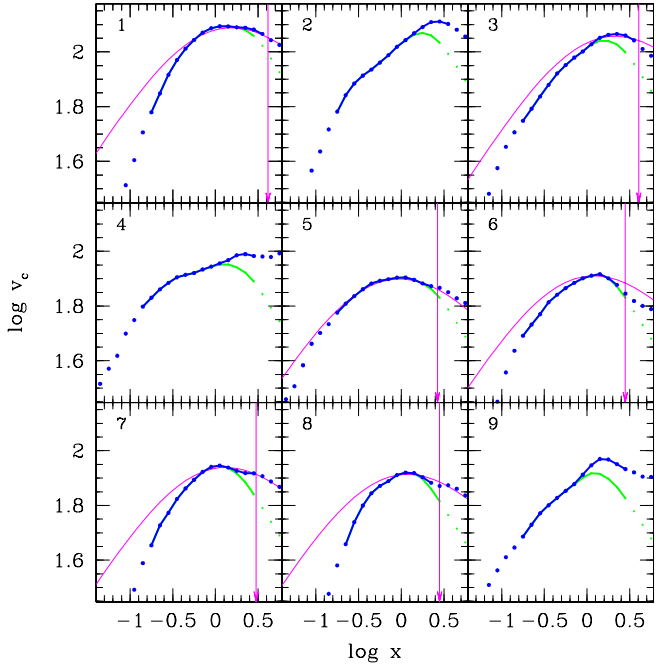,height=9cm}}
\centerline{\psfig{figure=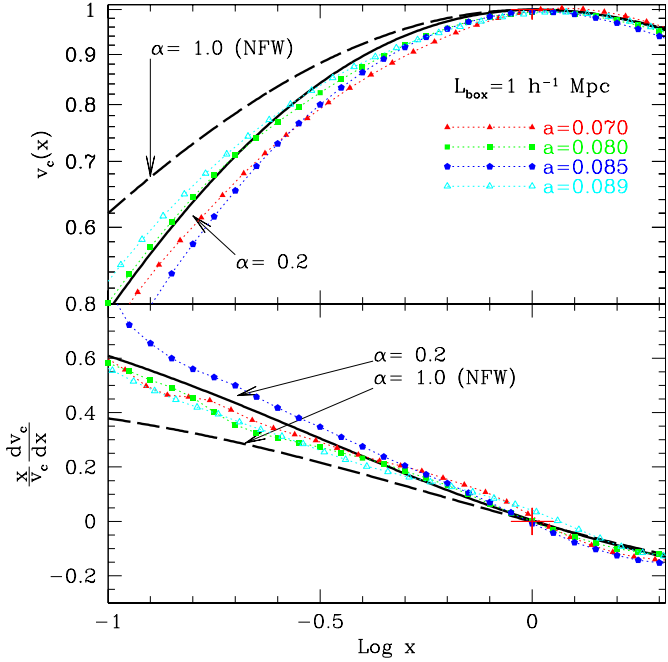,height=9cm}}
\caption{\label{fig:8pa1} {\em (Top)}. Dimensionless circular velocity,
  $v_c$, as a function of comoving radius, $x$, for the nine most
  massive halos in the 1 \Mpc simulation at $a=0.085$. The points are
  connected in the interval where the profile is reliable (the arrow
  indicates the virial radius). The smaller points are obtained using
  method (ii) (see text) and the thin solid line shows the NFW fit to
  the profile (for those profiles that are not undergoing major
  mergers). {\em (Bottom)}.  Normalised mean circular velocity $v_c$
  (top panel) and its slope (bottom panel) as a function of the
  normalised radius $x$ of the nine more massive halos in the
  simulation. The thick dashed line shows the NFW profile and the
  thick solid line a profile with $\alpha =0.2$. The mean profiles at
  at $a=0.070, 0.080, 0.085$ and $0.090$ are shown to ensure that the
  result is not affected by time dependent effects such as mergers and
  to increase the statistical significance of the result.}
\end{figure}

\begin{figure}
\centerline{\psfig{figure=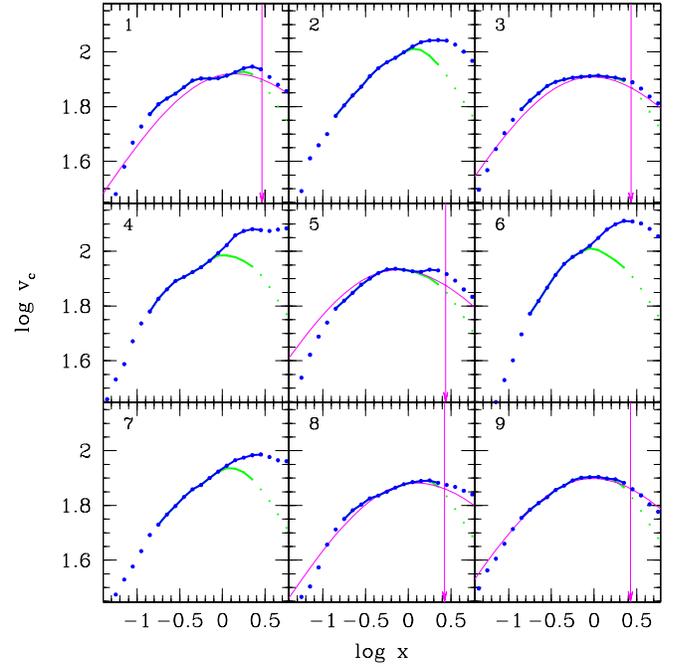,height=9cm}}
\centerline{\psfig{figure=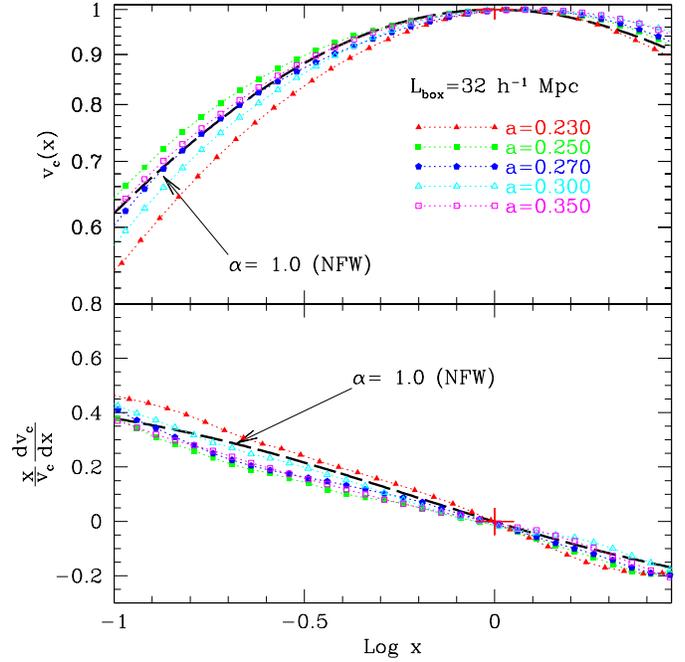,height=9cm}}
\caption{\label{fig:8pa2} {\em (Top)} Same as in Fig.~\ref{fig:8pa1} for the
  32 \Mpc box at $a=0.23$. {\em (Bottom)} Same as in Fig.~\ref{fig:8pa1} for
  the 32 \Mpc box at $a=0.23, 0.25, 0.27$ and $0.30$.}
\end{figure}

\begin{figure}
\centerline{\psfig{figure=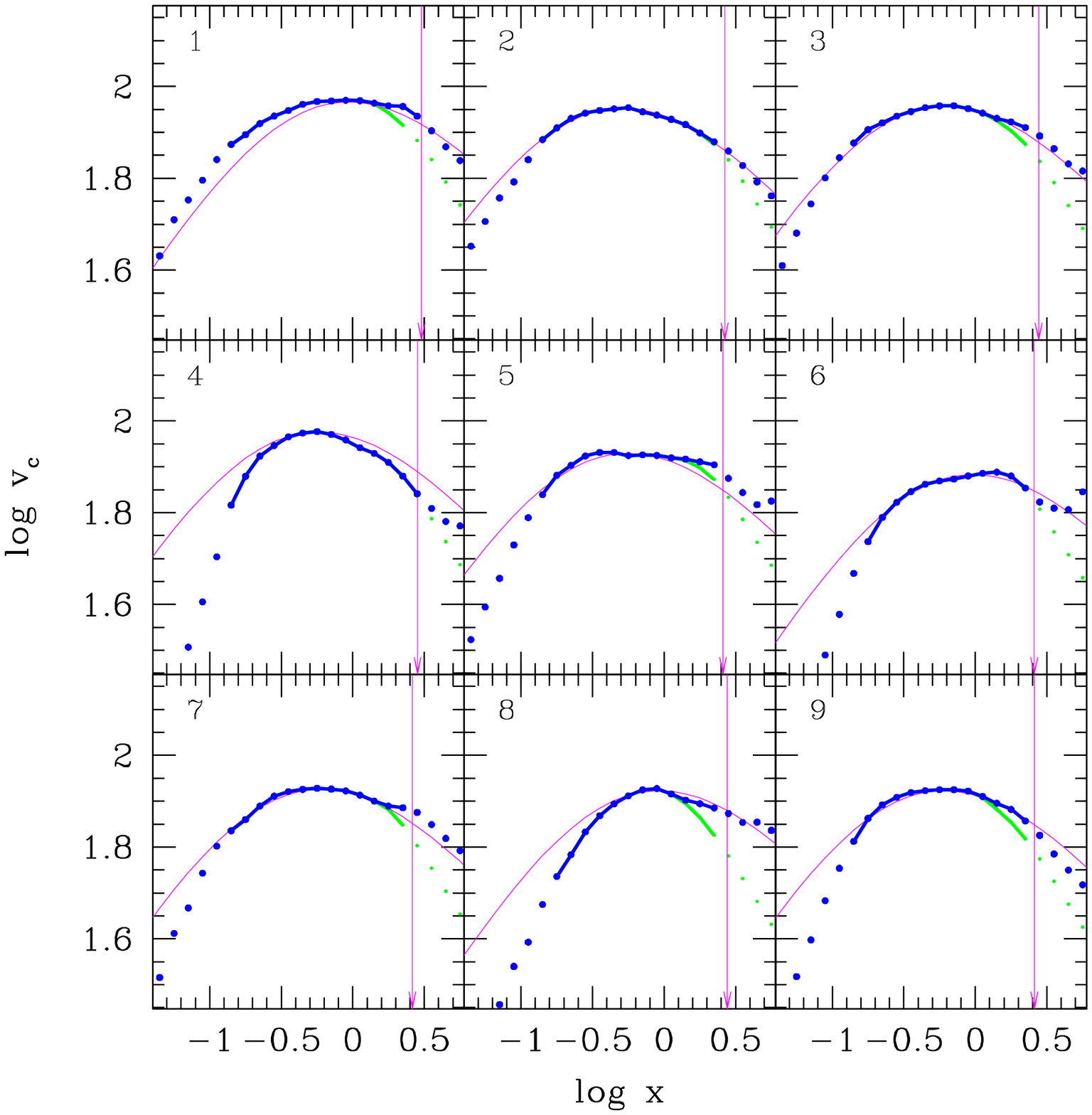,height=9cm}}
\centerline{\psfig{figure=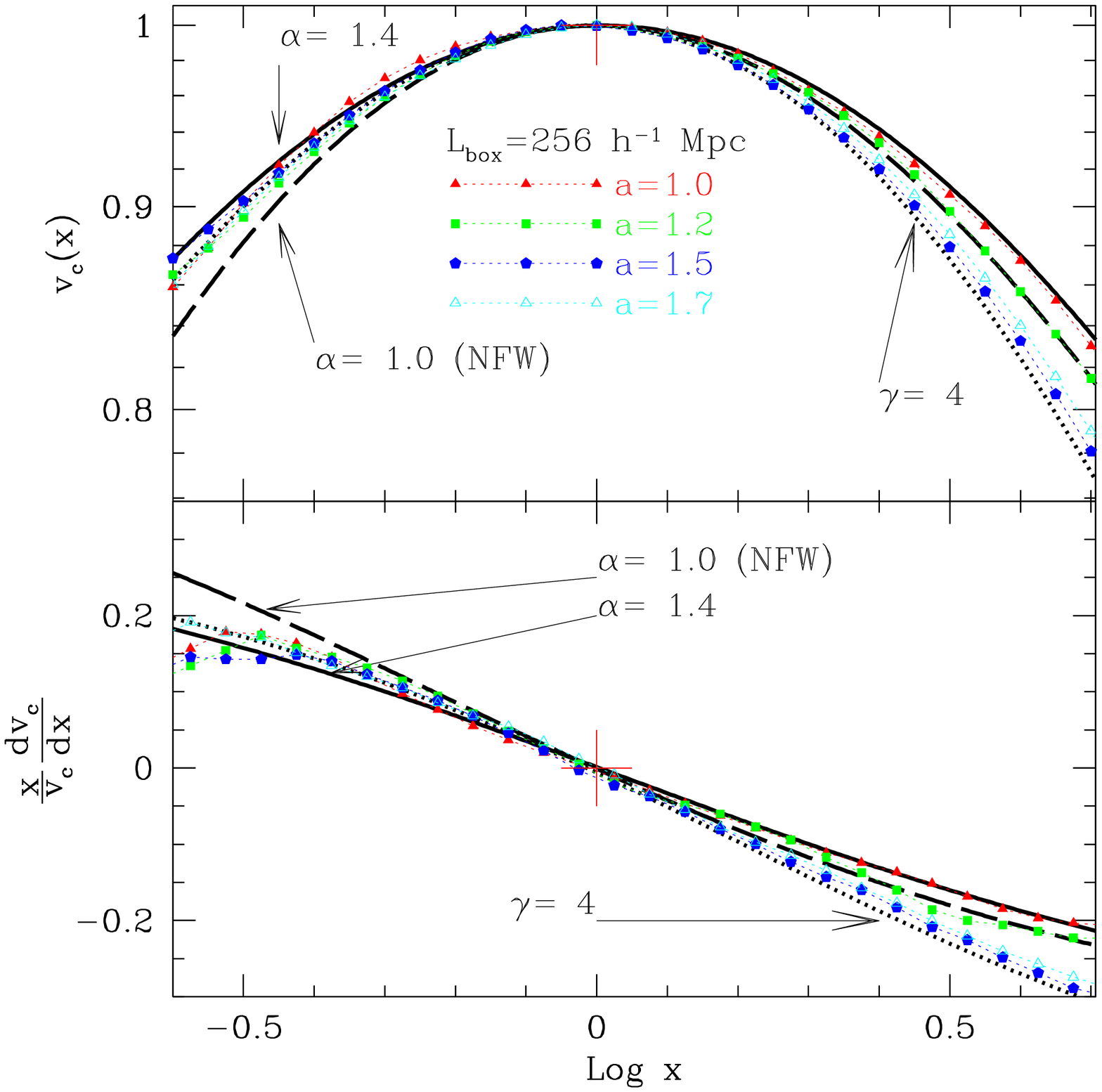,height=9cm}}
\caption{\label{fig:8pa3} {\em (Top)} Same as in Fig.~\ref{fig:8pa1} for the
  256 \Mpc simulation at $a=1.0$. {\em (Bottom)} Same as in
  Fig.~\ref{fig:8pa1} for the 256 \Mpc box at $a=1, 1.2, 1.5$ and
  $1.7$. The solid and dotted thick lines show two profiles both with
  $\alpha =1.4$, but with slopes of the outer profile $\gamma=3$ and
  $\gamma=4$, respectively.}
\end{figure}

In \figs~\ref{fig:8pa1}-\ref{fig:8pa3} ({\em top}) the dimensionless
circular velocities are shown for the nine most massive halos in the
simulations with box sizes $L_{box}=1, 32$, and 256 \Mpc,
respectively. The statistical errors, $\delta v_c$, on each point are
about the sizes of the symbols and therefore are not shown to avoid
excessive crowding. I show the profiles obtained using methods (ii)
(small dots) and (iii) (large dots). I do not show those obtained
using method (i) since they are almost indistinguishable from (ii). I
show the fit with a NFW profile obtained by matching the position and
value of the maximum of $v_c$ and the virial radius (arrow) for the
halos that are not severely perturbed by mergers.
The points are connected in the range of radii that are reliable:
$x_{mi} < x< x_\Delta$. Here $x_\Delta=2.37
(\Delta/178)^{-1/3}(N_p/10^4)^{1/3}$ is the dimensionless virial
  radius and $x_{mi}$ is the radius that contains a number of
  particles $N_p(x_{mi}) \ge 300(\rho(x_{mi})/10^4)^{1/2}$. This
  condition ensures that two-body relaxation is not important in
  flattening the inner profile (see \cite{Power:02} for details on
  convergence studies). Note that $x_{mi}$ is always larger than twice
  the Plummer softening length $2\epsilon=0.04$, therefore an
  incorrect force calculation does not affect the results at those
  radii.
  In summary, the sanity checks adopted are as follows.\\
1) Only radii well above twice the softening length are considered.\\
2) Only radii that enclose $N_p \ge 300(\rho/10^4)^{1/2}$ particles are considered.\\
3) No merging halos: profiles (ii) and (iii) give similar results for
the value and radius of the maximum circular velocity.\\
4) The same criteria are adopted for all the simulations with
different box sizes, $L_{box}$, and scale factor $a$.\\

\figs~\ref{fig:8pa1}-\ref{fig:8pa3} ({\em bottom}) show the mean
profiles obtained by scaling (shifting in logarithmic scale) the radii
to the same $x_{max}=1$ and normalising the maximum of $v_c$ to
$v_c^{max}=1$. Each radial bin, logarithmically spaced, is weighted
equally so long as $N_p \ge 300(\rho/10^4)^{1/2}$. Only the halos that
are not undergoing a major merger are considered. If all the ten
profiles or a different subset of profiles are considered, I find that
the resulting mean profile does not change significantly. Also, using
the arithmetic or geometric mean does not change the mean profiles. A
possible criticism to this method is that the halos analysed are not
isolated as in works where a single galaxy is re-simulated with higher
resolution.  Time-dependent effects could thus produce flatter or
steeper cores.  For this reason, I present the mean profile at
different redshifts (listed in the figure captions).  If the halos are
not relaxed to their final configuration the mean profiles should
differ at different redshifts.  Moreover, the increased statistical
sample of halos makes the result more robust.


In the 1 \Mpc simulation (see Fig.~\ref{fig:8pa1}) the slope of the
inner profiles is, on average, flatter (\ie, the circular velocity
rises more steeply) than the NFW profile. The typical masses of the
halos in this simulation are $M_{dm} \sim 10^8$ M$_\odot$. Even at
radii close to $v_{max}$ the circular velocity starts to deviate from
the NFW fit. The slope of the outer profile is instead consistent with
$\gamma =3$ as in the NFW formula.  Note that method (ii) does not
attribute several DM particles to the halo, producing an outer profile
that decreases more steeply. It is not clear if DENMAX fails not to
include those particles or if they effectively do not belong to the
halo. The definition of halo boundary (or virial radius) is somewhat
arbitrary and time dependent since halos are constantly growing.
Especially for small mass halos the flat mass function of the
accreting satellites and the short accretion timescale makes it more
difficult to define the outer edge of the galaxy. This is also true
for dwarf galaxies at $z=0$, even if the accreting satellites are
mostly invisible because they do not contain stars. For this reason an
``isolated'' dwarf galaxy cannot be found or defined in the same way
as for massive galaxies for which the mass function of accreting
satellites is dominated by halos with masses similar to the host halo.
There are halos for which the use of either method (iii) or (ii) gives
a different radius where the circular velocity is maximum.  Those
halos are excluded from the analysis since they are severely disturbed
by accreting satellites (but if they were to be included in the
analysis, the mean slope would remain flatter than $\alpha=1$). The
mean density profiles are best fitted by inner slopes in the range
$\alpha \sim 0.4 - 0.5$, equivalent to mean slopes of the inner
circular velocity $\beta \sim 0.75-0.8$.

In the 32 \Mpc simulation (see Fig.~\ref{fig:8pa2}) the NFW profile
provides a good fit to all the halos that are not severely disturbed
by accreting satellites. The typical masses of the halos in this
simulation are $M_{dm} \sim 5 \times 10^{12}$ M$_\odot$. The mean
density profiles are best fitted assuming inner slopes in the range
$\alpha \sim 0.9 - 1$, equivalent to mean slopes of the inner circular
velocity $\beta \sim 0.5 - 0.55$.

In the 256 \Mpc simulation (see Fig.~\ref{fig:8pa3}) the inner density
profile slopes appear slightly steeper than predicted by NFW (\ie,
flatter circular velocities).  Unfortunately, the number of particles
in the inner region is barely sufficient to start seeing a
discrepancy. A larger number of particles per halo is needed to
construct reliable rotation curves at small radii where the
discrepancy, expected by extrapolating the reliable part of the
rotation curve, should become evident.  The typical masses of the
halos in this simulation are $M_{dm} \sim 2 \times 10^{15}$ M$_\odot$.
The mean density profiles are best fitted by inner slopes in the range
$\alpha \sim 1.3 - 1.4$, equivalent to mean slopes of the inner
circular velocity $\beta \sim 0.3 - 0.35$. In \fig~\ref{fig:8pa3}
({\em bottom}), it appears that the slope of the outer profile becomes
steeper when $a >1$. For this reason, included in the figure are the
best fits for two cases: (i) when the outer slope of the profile is
$\gamma \approx 3$ at $a=1.0, 1.2$, and (ii) when the slope is $\gamma
\approx 4.0$ at $a=1.5, 1.7$.  This dependence of the outer profile
slope on the scale factor becomes more evident in the 1024 \Mpc
simulation and is discussed in the following paragraph.

\begin{figure}
\centerline{\psfig{figure=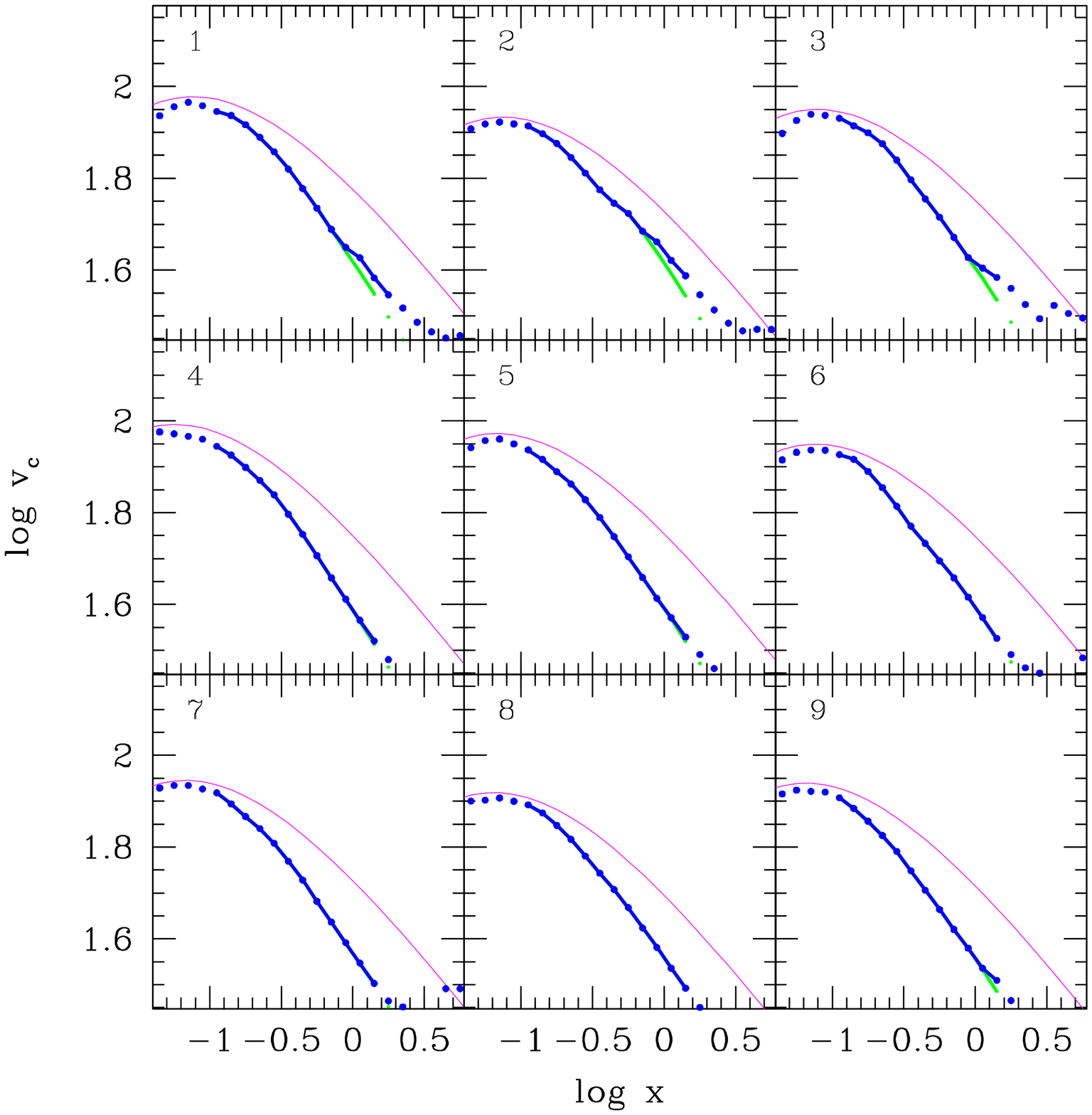,height=9cm}}
\centerline{\psfig{figure=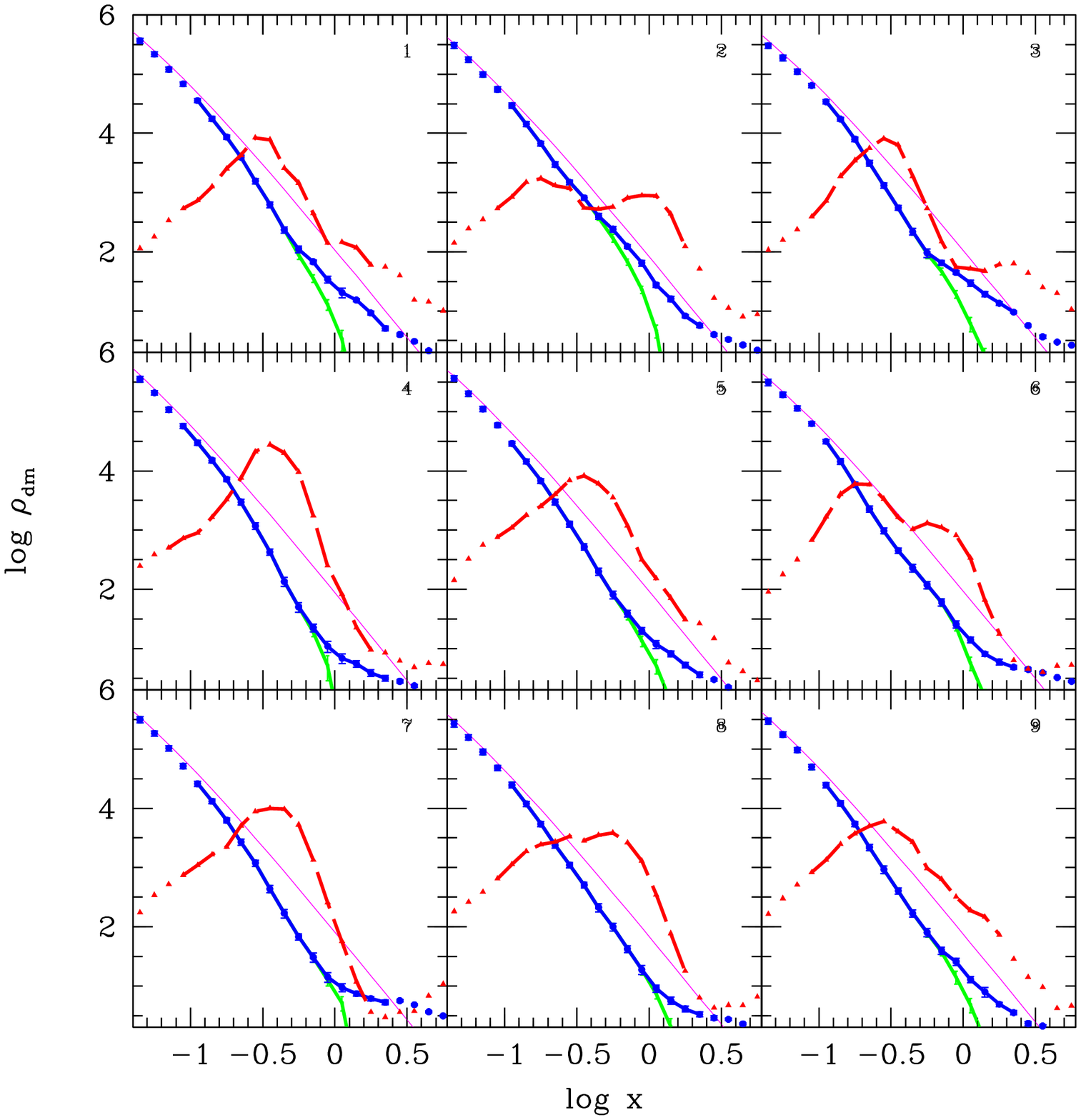,height=9cm}}
\caption{\label{fig:8pa4} {\em (Top)} Same as in Fig.~\ref{fig:8pa1} for the
  1024 \Mpc simulation at $a=10$. {\em (Bottom)} Density profile
  (solid lines) and slope of the density profile (dashed lines) for
  the 1024 \Mpc simulation at $a=10$. The thin solid line is the fit
  with the NFW profile. The outer slope of the density profile is
  $\gamma \approx 4$, steeper than in the NFW profile. This is
  probably caused by the accelerating universe.}
\end{figure}

In \fig~\ref{fig:8pa4} ({\em top}), the dimensionless circular
velocities are shown for the nine most massive halos in the simulation
with box size $L_{box}=1024$ \Mpc at $a=10$. The number of particles
in these halos is too small for measuring the slope of the inner
profile but it is evident that the outer slope is steeper than the NFW
profile.  \fig~\ref{fig:8pa4} ({\em bottom}) shows the density
profiles (solid lines) and slope of the density profiles (dashed
lines) for the nine more massive halos in the same simulation. The
outer density profile has a slope $\gamma \sim 4$. At $a =2.3$ the
slope is $\gamma \sim 3.5$ and increases to $\gamma \sim 4$ when the
scale factor is $a=4, 10, 100$ and 1000.  The physical explanation of
this result is likely to be related to the expansion rate of the
universe. When $\Omega_\Lambda(z) \sim 0$ at $a \simlt 0.5$ the
universe is decelerating and $\gamma \sim 3$. As $a$ increases to $a
\simgt 1$ the universe starts inflating (at $a=2$, $\Omega_\Lambda
\sim 0.95$ in a flat universe) and the slope of the outer profiles
becomes $\gamma \sim 4$. It is possible that even at $z=0$ the outer
profile of recently virialized clusters is slightly steeper than
$\gamma =3$.  According to the aforementioned interpretation, the
observation of a steep outer profile in nearby clusters would indicate
that the universe is accelerating (see also the results of
\cite{SubramanianCO:00} for low density universes).  In
\S~\ref{sec:prof} I show that for the halo profiles in this simulation
the radii, $r_{max}$, where the circular velocity is maximum, decrease
with increasing scale factor, consistent with the predictions assuming
an inner slope of the density profile $\alpha \sim 1.5$.

\subsection{Radiative Feedback Effects on the DM profile}\label{ssec:bary}

\begin{figure}
\centerline{\psfig{figure=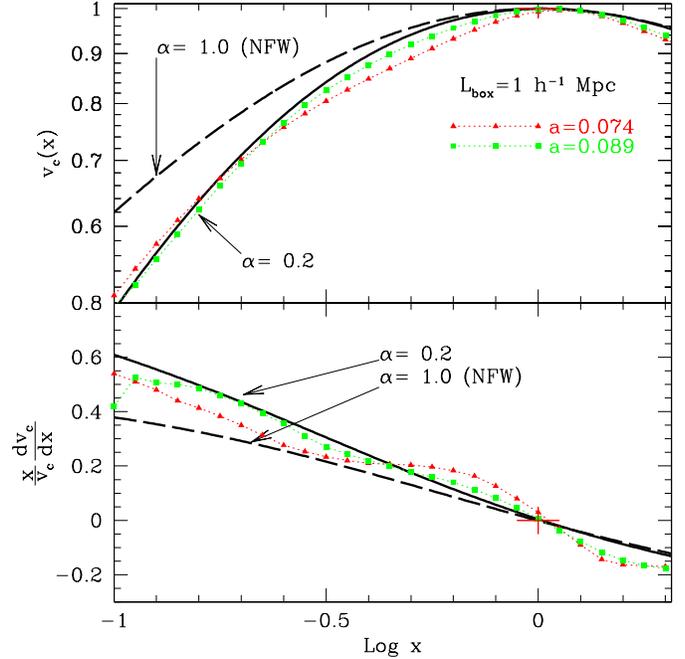,height=9cm}}
\caption{\label{fig:RGS} Same as in
  Fig.~\ref{fig:8pa1} {\em (Bottom)} for the 1 \Mpc simulation
  presented by Ricotti \etal (2002b) at $a=0.075$ and $0.090$ (see text).}
\end{figure}

\fig~\ref{fig:RGS} shows the mean $v_c$ for the nine most massive
halos in the simulation 256L1p3 presented by \cite{RicottiGSb:02} at
$a=0.075$ and $a=0.090$.  This simulation attempts to simulate
realistically the formation of the first galaxies in the universe
modeling radiative feedback effects in the early universe.
The parameters of the simulation are the same as in the 1 \Mpc
simulation presented in this work. But in addition to DM particles it
includes gas dynamics, star formation using a Schmidt-Law, metal
enrichment from star formation and radiative transfer. Radiative
feedback effects produced by UV radiation emitted by the first stars
are calculated following the molecular processes involving H$_2$
formation, destruction and cooling. From the results presented in this
work it appears that radiative feedback processes are not responsible
for the flattening of the DM profile in small-halos. The slopes of the
inner profile of the halos in this simulation are similar to the
corresponding N-body simulation, or perhaps slightly steeper.

\section{Constant concentration DM profiles}\label{sec:prof}

\begin{figure}
\centerline{\psfig{figure=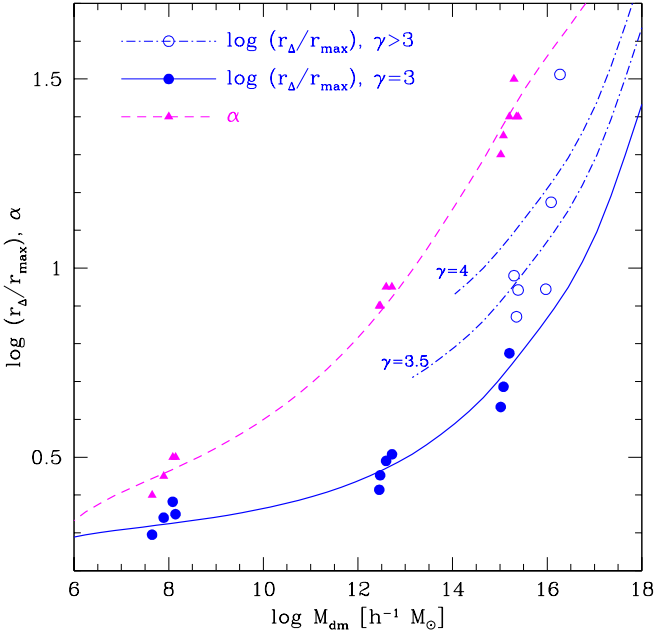,width=9cm}}
\caption{\label{fig:cd} $r_\Delta/r_{max}$ (circles) and
  $\alpha$ (triangles) for the simulations listed in Table~\ref{tab:one}
  as a function of the halo DM mass $M_{dm}$. The open circles show
  $r_\Delta/r_{max}$ when $\gamma>3$ at $a>1$. The dashed line shows the
  theoretical value of $\alpha$ (see text).  The solid and dash-dotted
  lines show $r_\Delta/r_{max}$ for the profile given in Eq.~(\ref{eq:g})
  using $\gamma=3$ and $\gamma=4$, respectively.}
\end{figure}

In comoving coordinates, $r$, all of the simulated halos in this
work can be fitted by a profile with the form \citep[\cf,][]{Zhao:96}
\begin{equation}
\rho_{dm}(X) \propto {1 \over X^\alpha(1+X)^{\gamma-\alpha}},
\label{eq:g}
\end{equation}
where $X=c_\Delta r/r_\Delta$, $r_\Delta$ is the comoving virial
radius, $\alpha=\alpha(M_{dm})$ is a function of the halo mass,
$M_{dm}$, and $c_\Delta \sim 7$ is the concentration parameter that,
as I will show, is a constant. In physical coordinates, $R$, it is
simply $X=c_\Delta R/R_\Delta$, where
$R_\Delta=(1+z)^{-1}r_\Delta$ is the virial radius.  Given the initial
power spectrum of density perturbations, the parameter $\alpha$ is a
function of the halo mass only. The values found in this work agree
with the predictions of \cite{SubramanianCO:00} (see
Fig.~\ref{fig:cd}). An acceptable fit to the theoretical value of
$\alpha$ for the case of the $\Lambda$CDM cosmology, adopted in this
work, has the form
$\alpha \approx 1.3+(M_{dm,14}^{1/6}-1)/(M_{dm,14}^{1/6}+1)$, where
$M_{dm,14}=M_{dm}/(3 \times 10^{14}$ M$_\odot$). This fit is five
percent accurate for masses $10^6 \simlt M_{dm}/M_\odot \simlt
10^{18}$.

The logarithmic slope, $\gamma$, of the outer parts of the halo
appears to depend on the acceleration of the universe: when the scale
parameter is $a \simlt 1$, the slope is $\gamma \approx 3$ as in the
NFW profile, but is $\gamma \approx 4$ at $a >1$ when $\Omega_\Lambda
\sim 1$ and the universe is inflating. Equation~(\ref{eq:g}) is
normalised imposing the condition that the integrated mass inside the
virial radius is $m_{dm}(X=c_\Delta)=M_{dm}$ (see appendix~\ref{app}).
The comoving virial radius is defined as
\[
r_\Delta \equiv (32~{\rm kpc})\left({\Delta \over 178}\right) \left(M_{dm}
\over 10^6 M_\odot \right)^{1/3}, 
\]
to satisfy the relation $M_{dm} =(4\pi/3) \rho_0\Delta r_\Delta$,
where $\Delta$ is the mean overdensity of the halo and $\rho_0$ is the
mean DM density at $z=0$.  $\Delta$ depends on the cosmology since it
is related to the mean overdensity and the time it takes for a density
perturbation to stop expanding with the universe and ``turn around,''
starting the collapse.  According to the simple top-hat spherical
collapse approximation, $\Delta \approx 178\Omega_0(z)^{0.45}$ for a
flat universe with a cosmological constant \citep{Eke:98}.

I now explain why this profile has the property of a constant
concentration parameter.  The changing slopes of the inner and outer
parts of the profile mimics the variation of the concentration
parameter in the NFW profile, shifting the value of the radius,
$r_{max}$, where the circular velocity reaches its maximum value. For
the NFW profile (\ie, for $\alpha=1$)
$r_\Delta/r_{max}=c_\Delta^{NFW}/2.16$, therefore the location of
$r_{max}$ with respect to the virial radius is inversely proportional
to the concentration parameter. In the general case
\begin{equation}
{r_\Delta \over r_{max}}= {c_\Delta \over f(\alpha, \gamma)}.
\label{eq:ff}
\end{equation}
The location of the maximum circular velocity depends on the inner and
outer profile slopes and on the concentration parameter.  In
appendix~\ref{app} I show that a good approximation for the function
$f(\alpha, \gamma)$ is $f \approx (4 / \gamma)^{2.68}(2-\alpha)$.  In
\fig~\ref{fig:cd} it can be seen that the values of $r_\Delta/r_{max}$
(circles), for all the simulations in this work (listed in
Table~\ref{tab:one}), agree with \eq~(\ref{eq:ff}), assuming constant
concentration parameter $c_\Delta=7$. Moreover the values of $\alpha$
(triangles) are in good agreement with the theoretical relationship proposed
by \cite{SubramanianCO:00}, $\alpha=(9+3n)/(5+n)$ (dashed line),
where $n$ is the effective slope of the power spectrum. The solid and open
circles show $r_\Delta/r_{max}$ measured in profiles with $\gamma=3$
(\ie, at $a<1.2$) and with $\gamma>3$, respectively.  The solid and
dash-dotted lines show $r_\Delta/r_{max}$ given by \eq~(\ref{eq:ff})
using the theoretical value of $\alpha$ (shown by the dashed line),
$\gamma=3$ and $\gamma=3.5-4$, respectively. In the 1024 \Mpc box
simulation I find that $\gamma \sim 4$. The inner slope of the
profile cannot be measured because of the small number of particles in
the inner regions. But the measured values of $r_\Delta/r_{max}$ are
consistent with values of the inner slope $\alpha \sim 1.5$,
predicted by the theory.

\section{Discussion}\label{sec:disc}

The results presented in this work and their interpretation seem to
disagree with most previously published work, although the simulations
presented here are not of substantially higher resolution than those
in the literature.  In this section I analyse the discrepancies with
previous works and point out that the results do not disagree strongly
with most previous works, save for the 1 \Mpc simulation. The result
of the 1 \Mpc simulation cannot be compared with previous works since
a similar simulation has not been performed before. The joint analysis
of the three simulations with $L_{box}=1, 32$ and 256 \Mpc suggests
that the slope of the inner part of the DM profile depends on the
power spectrum of linear perturbation. This interpretation of the
results disagrees with the common view of universality of DM profiles,
although several groups have found results that contrast with this
idea, both in the context of galactic dynamics
\citep[][\eg,]{NipotiLC:03} and cosmology \citep{SyerW:98,
  Kravtsov:98, JingS:00, Jing:00}. Here I identify and summarise
three possible reasons why the present results differ from
a large number of published work on N-body simulations.

\begin{enumerate}
\item {\em The profiles of halos with typical masses resolved by the 1
    \Mpc simulation have not been analysed before.}  This simulation
  is equivalent to a scale free simulation with linear power spectrum
  $P(k) \propto k^n$ with $n =-2.7$ and Einstein-de Sitter cosmology,
  since at $z > 2$ the cosmological constant contribution is
  negligible.  Previous works have studied scale free simulations with
  power spectrum index $n \ge -2$. For instance \cite{Cole:96} studied
  the $n=0, -1, -2$ cases and \cite{Navarro:97} the $n= 0, -0.5, -1,
  -1.5$ cases. The simulations in those works have the same resolution
  as in the present study.  Their results do not contrast with the
  findings of this work because when $n \simgt -2$ the inner slope of
  DM profiles is also $\alpha \approx 1-1.5$. But, as mentioned, the
  resolution of the simulations (about $10^4-10^5$ particles per halo)
  is marginally sufficient to discern between $\alpha \sim 1$ and
  $\alpha \sim 1.5$. The $n=-2.7$ case is markedly different from the
  others since it corresponds to a scenario in which all mass
  perturbations become non-linear at about the same time (an index $n
  < -3$ would correspond to the top-down formation scenario, in which
  small mass halos form after the large ones). For this reason, I
  believe that the result found analysing the 1 \Mpc simulation is new
  and complements previous studies.
    
\item {\em Interpretation of the results.}  Theoretically it is
  unclear what determines the slope of the inner DM profiles. Some
  works argue that the slope depends on the power spectrum of initial
  perturbations \citep{SyerW:98,SubramanianCO:00}, others argue
  that the slope tends asymptotically to a steep slope $\alpha \sim
  1-2$ \citep{AlvarezS:03, Dekel:03}.  Recent still unpublished
  high-resolution N-body results from Navarro and collaborators show
  that the inner slope does not converge to $\alpha=1$ but the slope
  slowly flattens to $\alpha \simlt 1$. This result contrasts with most
  aforementioned semi-analytical results indicating that physical
  processes that determine the density profile of DM halos and their
  time evolution are not well understood.
  
  The halos analysed in the present work are the most massive halos at
  a given redshift. For this reason they are recently virialized halos
  with different masses, compared after about the same number
  dynamical times from their formation. The halos analysed in
  high-resolution N-body studies are usually chosen to be
  ``isolated'': halos that did not experience ``recent'' major
  mergers. Therefore, depending on their mass, their age from the
  virialization differs greatly at $z=0$. Semi-analytical works study
  the equilibrium profile of halos given an accretion history. The
  different evolutionary states of the halos studied in literature
  could explain the discrepancies in the results if the time scale to
  reach the equilibrium configuration is long compared to the
  accretion time. I find that the halo profile depends on the
  accretion history. But it is possible that this is true only during
  a transient period of time, before the halo relaxes to a different
  equilibrium configuration.  This concern is especially valid for the
  1 \Mpc simulation since in a scale free simulation with $n \sim
  -2.7$ the halos continuously accrete satellites. The result is
  therefore statistically significant at redshift about 10 but the
  shallow slope of the profiles might not survive to redshift $z=0$.
  Even if the shallow profiles are transient, depending on the time
  scale to reach the equilibrium configuration, the result at $z=10$
  might suggest a reason for why not all dwarf galaxy halos at $z=0$
  present shallow cores.  According to the work of \cite{Dekel:03},
  substructure is not tidally striped but instead is compressed when
  the halo profile is shallow. Therefore substructure could survive
  undigested for quite a long time.  Moreover even at $z=0$, small
  mass halos, due to their larger number density, are expected to
  capture dark satellites of non-negligible mass with higher frequency
  than larger mass halos.  Further work is needed, and is currently in
  progress \citep{RicottiW:03}, to extend the result of the present
  work from redshift $z=10$ to redshift $z=0$.  The aim of this study
  is to understand if the flat profiles at $z=10$ are in a stable
  configuration that would survive for a Hubble time and if the
  profiles are consistent with the observed flat cores of dSph
  galaxies in the local group. Preliminary results show that the
  dynamical state of the halos is stable and the inner profiles remain
  flat for $20$ Grys if the halo is evolved in isolation.
  
  To further support the case for a dependence of the inner profile on
  the mass of the halos I show that flat inner profiles are associated
  with an almost isotropic velocity dispersion of DM particles while
  steeper profiles have a velocity dispersion slightly biased in the
  radial direction. This result is shown in Fig.~\ref{fig:beta}. The
  lines show the parameters $\beta=1-\sigma^2_\theta/\sigma^2_r$
  (dashed lines) and $\beta^*=1-\langle v_\theta^2 \rangle/\langle
  v_r^2 \rangle$ (solid lines) as a function of the scaled radius
  $x=r/r_{max}$, where $r_{max}$ is the radius where the circular
  velocity is maximum.  Here, $\sigma^2_\theta=\langle v_\theta^2
  \rangle-\langle v_\theta \rangle^2$ and $\sigma^2_r=\langle v_r^2
  \rangle-\langle v_r \rangle^2$, where $v_r$ is the radial velocity
  and $v_\theta$ is the tangential velocity. The $\beta$ parameter
  measures the degree of anisotropy of the velocity dispersion:
  $\beta=0$ corresponds to an isotropic velocity dispersion, $\beta<0$
  to a tangentially biased velocity dispersion and $0<\beta<1$ to a
  radially biased velocity dispersion.  The lines from top to bottom
  show the mean value of $\beta$ and $\beta^*$ for the 9 most massive
  halos in the 256 \Mpc, 32 \Mpc and 1 \Mpc simulations. The shallow
  inner DM profiles in the 1 \Mpc simulation have almost isotropic
  velocity dispersion.  Steeper DM profiles in the 32 \Mpc and 256
  \Mpc simulations have a velocity dispersion slightly biased in the
  radial direction. This result is consistent with the analytical
  expectations based on the solution of the Jeans equations in
  spherical systems.

\begin{figure}
\centerline{\psfig{figure=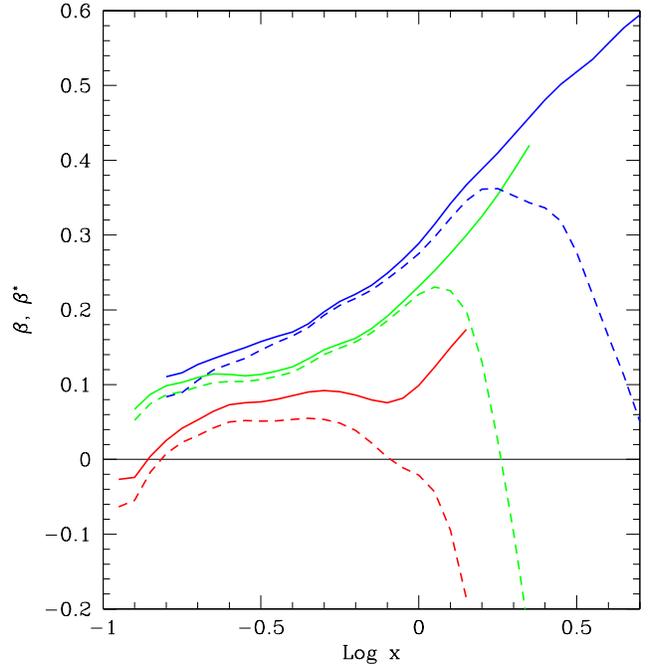,width=9cm}}
\caption{\label{fig:beta} $\beta=1-\sigma_\theta^2/\sigma_r^2$ (dashed
  line) and $\beta^*=1-\langle v_\theta^2 \rangle/\langle v_r^2
  \rangle$ (solid lines) as a function of the scaled radius
  $x=r/r_{max}$.  The lines from top to bottom show the mean value of
  $\beta$ and $\beta^*$ for the 9 most massive halos in the 256 \Mpc,
  32 \Mpc and 1 \Mpc simulations (shown in the top panels of
  Figs.~3-5).  The shallow inner DM profiles in the 1 \Mpc simulation
  have almost isotropic velocity dispersion ($\beta \sim 0$). Steeper
  DM profiles in the 32 \Mpc and 256 \Mpc simulations have a velocity
  dispersion slightly biased in the radial direction (\ie, $\sigma_r >
  \sigma_{\theta}$).}
\end{figure}

\item {\em Method and analysis of the results} In this paper I present
  statistical results without preselecting the halos to re-simulate
  with higher resolution as has been done in previous studies. In
  order to compare simulations with observations a statistically
  significant sample should be analysed.  Halos that did not reach an
  asymptotic equilibrium configuration (if they exist) or halos
  presenting substructures produced by recent mergers should be
  included in the analysis if they appear to be statistically
  significant.  Especially if there is not an easy way to determine the
  dynamical stage of a halo because the accreting substructure is
  invisible (small halos at $z=0$ should accrete mostly dark halos),
  understanding what fraction of dark halos of a given mass is in a
  transient configuration is relevant to solve the problem of the
  DM cores observed in a fraction of galaxies at low redshift.
  
  As expected the halos studied in this work show substantial
  substructure. 3D images of the halos show that numerous satellites
  are present, independently of the halo mass.  Perhaps in small mass
  halos substructure can survive ``undigested'' closer to the halo
  cores.  Nevertheless, in most cases, this does not affect the
  approximately spherical symmetry of the halos and the
  determination of the halo centres.  The halo centres are calculated,
  using DENMAX, as the minimum of the gravitational potential. I also
  calculate the centres as the maximum of halo density and as the
  centre of mass of bound particles. The halo centres calculated using
  those alternative definitions are practically coincident and choosing
  either of them does not change the profiles. Exceptions are
  profile 8 in the 1 \Mpc simulation and profile 4 in the 256 \Mpc
  simulation that have maximum density centres offset by
  2$\epsilon$, where $\epsilon$ is the Plummer softening parameter.
  
  In this work the inner slope of the profiles and the core radii are
  determined fitting the circular velocities instead of the density
  profiles.  This procedure partially remove the degeneracy between
  the slope $\alpha$ of the inner profile and the core radius $r_s$.
  This is illustrated in Fig.~\ref{fig:deg}, which shows the maximum
  likelihood estimate of the parameters $\alpha$ and $r_s$ fitting the
  density profile (left) or the circular velocity (right).  The
  $1\sigma, 2\sigma$ and $3\sigma$ confidence contours are shown
  assuming constant relative errors on the density and circular
  velocity in each logarithmically spaced radial bin.
\end{enumerate}
\begin{figure*}
\centerline{\psfig{figure=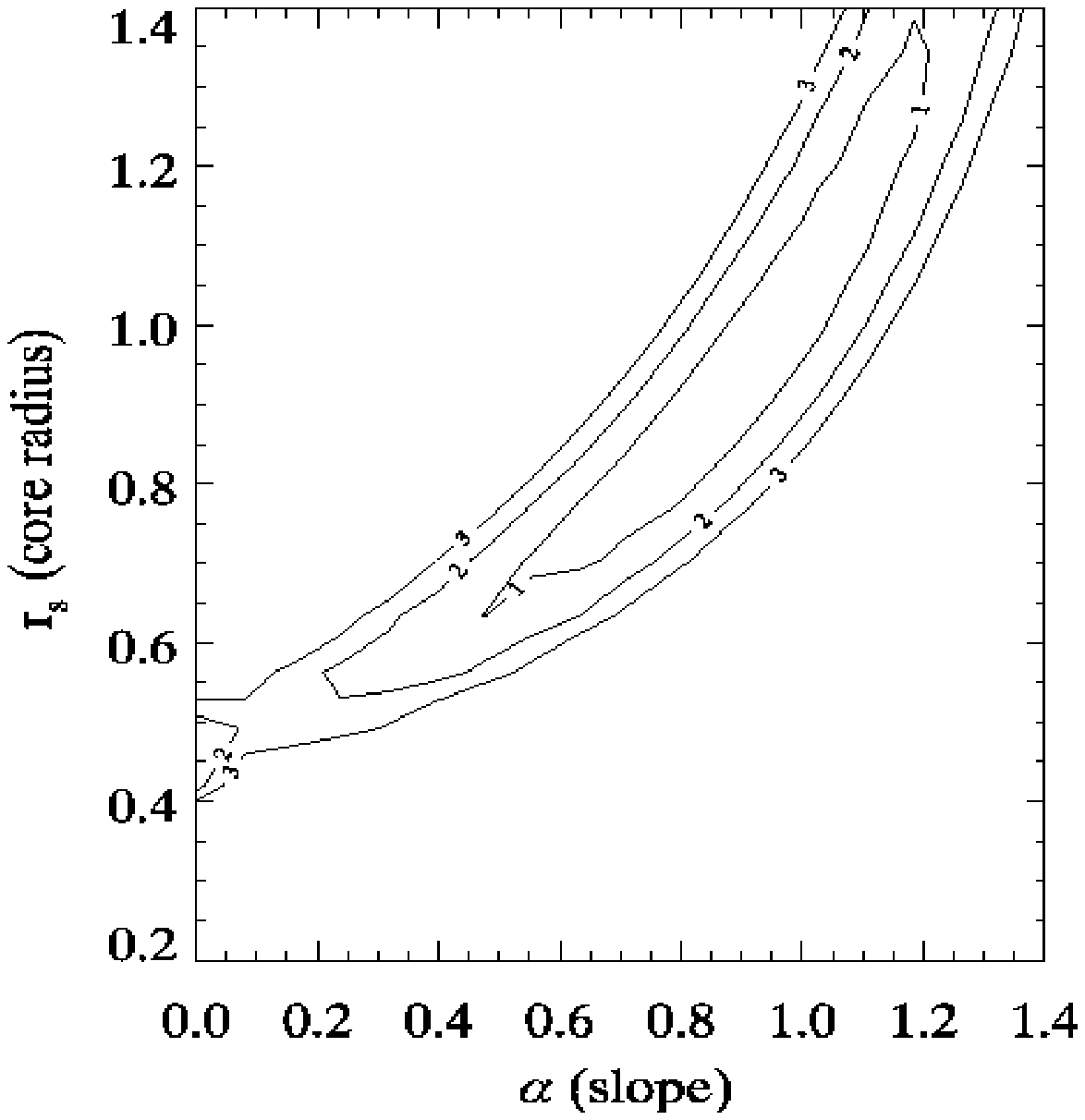,width=9cm}
\psfig{figure=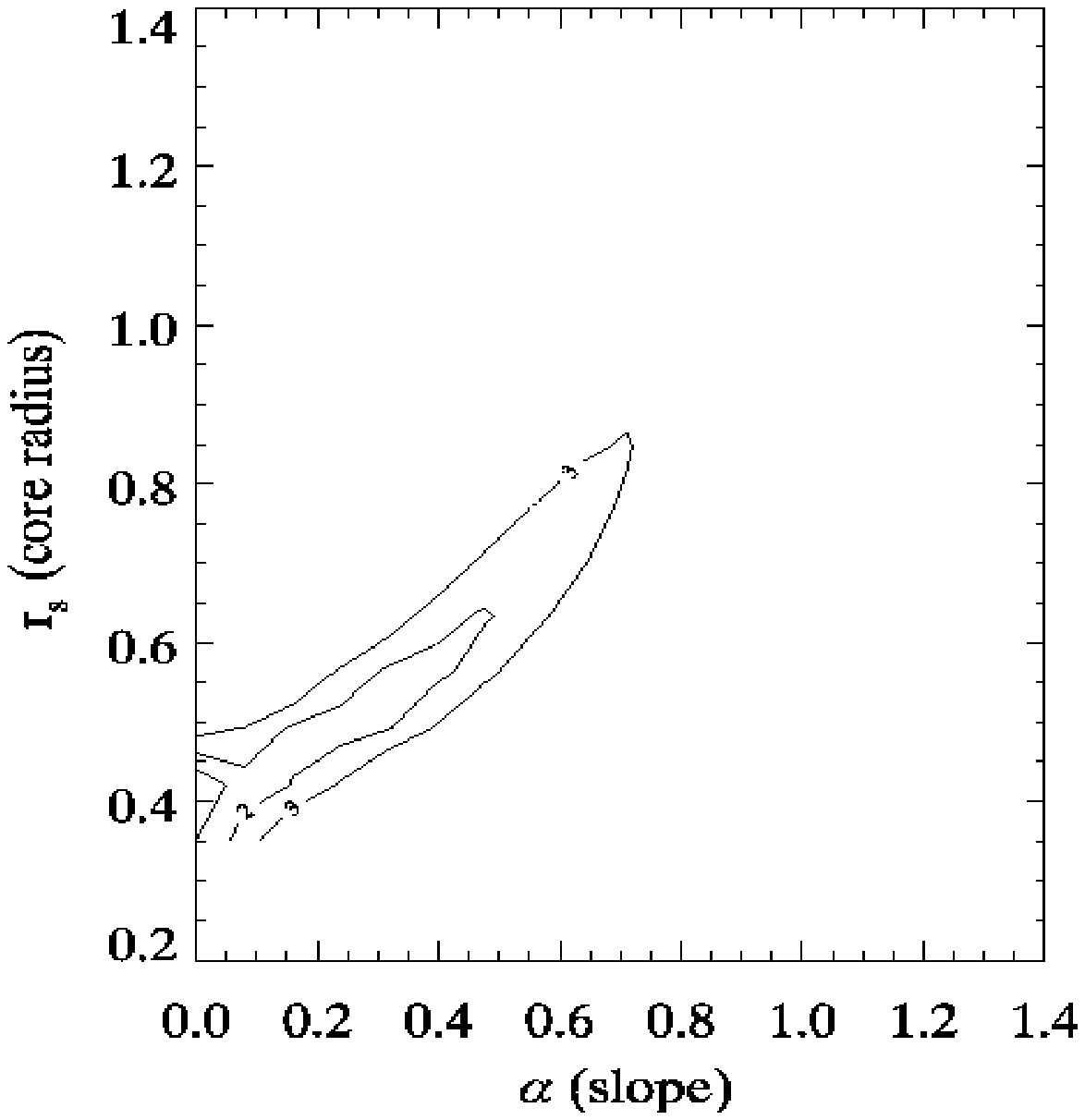,width=9cm}}
\caption{\label{fig:deg}  Maximum likelihood estimate of the core
  radius, $r_s$, and inner slope, $\alpha$.  The $1\sigma, 2\sigma$
  and $3\sigma$ confidence contours are shown. The degeneracy between
  the core radius and inner slope when the density profile is fitted
  with the generalised profile given in Eq.~(\ref{eq:g}) ({\em left
    panel}) can be partially broken fitting the circular velocity
  ({\em right panel}).}
\end{figure*}

\section{Conclusions}\label{sec:con}

Two main problems faced by CDM cosmologies are related to the
properties of small mass galaxies. Namely, (i) the number of visible
galactic satellites (dwarf galaxies) in the Local Group is smaller
than predicted by N-body simulations \citep{Moore:99, Klypin:99} and
(ii) the flatness of the DM cores in dwarf and LSB galaxies is not
reproduced by N-body simulations. The first problem can be solved by
including radiative feedback effects \citep{Chiu:01, RicottiGSb:02} in
cosmological simulations. Galaxies with masses $M_{dm} \simlt
10^8-10^9$ M$_\odot$ are too small to retain the gas that is
photoevaporated by stars or by the ionising background after
reionization. Consequently, the luminosity of most dwarf galaxies is
predicted to be too low to be detected (\ie, the mass-to-light ratio
should increase for halos with smaller masses) or zero for very small
mass halos. Here I have shown results of N-body simulations suggesting
that the second problem might not contradict the predictions of
CDM cosmologies. But further investigation is needed to understand if
the dwarf galaxies showing shallow DM profiles at $z \sim 10$ (\eg,
just after the time when most of the dwarf galaxies formed) are in
a configuration that will survive unchanged until redshift $z=0$.
In summary the main findings of this work are:
\begin{itemize}
\item The slope, $\alpha$ of the inner profile of DM halos is
  determined by the mass function of the accreting substructure (\ie,
  the initial power spectrum of initial perturbations). Dwarf galaxies
  have on average flatter DM cores, and clusters, steeper cores than
  galaxies similar to the Milky Way for which the
  NFW profile is a good fit to the halo density profile.

\item The logarithmic slope, $\gamma$, of the outer parts of the halo
  appears to depend on the acceleration of the universe: when the
  scale parameter is $a \simlt 1$, $\gamma \approx 3$ as in the NFW
  profile, but $\gamma \approx 4$ at $a >1$ when $\Omega_\Lambda
  \sim 1$ and the universe is inflating.
  
\item A density profile in the form $\rho_{dm} \propto
  X^{-\alpha}(1+X)^{-\gamma+\alpha}$, where $X=c_\Delta r/r_\Delta$,
  $\alpha$ is a function of the halo mass ($\alpha \approx
  1.3+[M_{dm,14}^{1/6}-1]/[M_{dm,14}^{1/6}+1]$, where
  $M_{dm,14}=M_{dm}/[3 \times 10^{14}$ M$_\odot$]) and $\gamma \sim
  3$ (but $\gamma \sim 4$ if $\Omega_\Lambda \sim 1$), can fit all the
  profiles from dwarfs to superclusters with constant concentration
  parameter $c_\Delta \simeq 7$.  This provides a physical explanation
  for the variation of the concentration parameter in the NFW profile
  as a function of the halo mass: the changing slopes mimic the
  variation of the concentration parameter by shifting the radius
  where the circular velocity reaches its maximum value.
  
\item The dependence of $\alpha$ on the mass of the halo, and
  therefore on the slope of the power spectrum, $n$, agrees with the
  theoretical expectation $\alpha = (9+3n)/(5+n)$, derived assuming
  that ``undigested'' satellites determine the halo structure in the
  inner regions \citep{SubramanianCO:00}. Note that if the DM halo
  develops a flat core, the chance for accreting satellites to
  survive tidal disruption is larger \citep{Dekel:03}.
\end{itemize}

The shape of the profiles for a given mass and redshift present
significant statistical variations. This can be attributed to two
factors. Firstly, galaxies are not isolated entities and especially at
high redshift are severely disturbed by neighbouring satellites or by
undergoing major mergers. This produces irregularities in the
azimuthally averaged density profile. Secondly, the accretion history
of satellites has statistical variation and depends on the local
environment. A dwarf galaxy forming at high redshift from a large
$\sigma$ peak of the power spectrum of initial perturbations (the
type of galaxy simulated in this work) could have an inner
profile different from the same mass galaxy forming at lower redshift
from a smaller $\sigma$ perturbation. Also, the slope of the profile
could differ if a galaxy forms in a void or in an overdense region,
even for galaxies with similar masses and radii.

I have identified three main reasons to explain why previous studies
\citep[\eg,][]{Eke:01, Moore:99} found that the inner slope of the DM
profiles are not determined by the power spectrum or accretion history
(but see also \cite{SyerW:98, Kravtsov:98, JingS:00, Jing:00} for a
different point of view). (i) The profiles of halos with typical
masses resolved by the 1 \Mpc simulation have not been analysed before.
(ii) The dynamical state of the the halos at $z=10$ might be different
at $z=0$. But there are not published high-resolution simulations for
dwarf-size halos ($M_{DM} \sim 10^8$ M$_\odot$) at $z=0$. The classic
results of \cite{Cole:96, Navarro:97} are based on the same type of
simulations presented here. (iii) Most of the computational efforts
have focused on simulating a single galaxy-size or cluster-size halo
with mass $M_{dm} \simgt 10^{12}$ M$_\odot$ at $z=0$ with higher
resolution (about $10^6$ particles per halo) than in this study. Such
a large number of particles is required to study the profile in the
very inner regions of the halo. But the task of achieving reliable
profiles with such a high spatial resolution is sensitive to numerical
integration errors and requires careful resolution studies
\citep{Power:02}; perhaps this is a reason for the disagreement
between groups, using different codes, on the slope of the inner
profile.  In this work I find that the shape of the DM profiles
present significant scatter around the mean. It is, therefore,
important to analyse a significant statistical sample of halos in
order to measure the mean profile.  Works where a single halo is
re-simulated with higher resolution are affected by a bias, difficult
to control, determined by the criteria for picking the halos to
re-simulate. Finally, the fitting method could be important as well
for the results. The degeneracy between the slope of the inner profile
and the value of the core radius is more easily broken fitting the
circular velocities instead of the density profiles. This is because
the radius where the circular velocity is maximum is univocally
determined by the value of the core radius, $r_s$ (for the NFW profile
is $r_{max}=2.16 r_s$). The halo profiles at any redshift and mass,
when renormalised matching the values and radii of the maximum
circular velocities, have to be identical if their profiles can be all
fitted by the NFW profile.

\cite{Moore:99} find steep profiles in cluster size halos ($\alpha
\sim 1.5$), in agreement with the results of this work. The results of
this work also agree with the classic NFW result for Milky Way size
galaxies but not with \cite{Moore:99} for halos with the same mass.
In this work the spatial resolution is not as large as in the
aforementioned studies, therefore I cannot study the slope of the DM
profile in the very centre of the halos. Nevertheless, I find evidence
for disagreement with the NFW predictions even at relatively large
radii. Perhaps this new result has been found because it is the first
time that a simulation of halos with masses typical of dwarf galaxies
and mass resolution $M_p \sim 10^3$ M$_\odot$ has been performed. The
result for the cluster size halos could also be understood with
similar reasoning (\ie, the cluster masses are larger than in previous
studies: $M_{dm} \sim 10^{15}-10^{16}$ M$_\odot$).  The steeper core
density profile found in clusters may not be as evident as the flatter
core found in dwarf galaxies but, combined with the result for normal
galaxies, the findings from the three simulations strengthen the case
for an inner slope that changes with the mass of the halos.

I conclude this work with two final remarks:
\begin{itemize}
\item[-] According to the scaling relation $\alpha = (9+3n)/(5+n)$,
  the inner slope of DM profile of dwarf galaxies is more sensitive to
  the power spectrum index, $n$, than more massive galaxies:
  $\delta\alpha = 1.5 \delta n$ if $n \sim -3$. A tilt $\delta n
  \approx 0.2$ in the initial power spectrum should produce a
  variation $\delta \alpha \approx 0.3$ of the inner slope of the DM profile.
  
\item[-] The best place to look for the effects of the cosmological
  constant is the outer profile of clusters and superclusters at low
  redshift. The prediction is for a slope steeper than $\gamma =3$.
  Perhaps gravitational lensing studies could be useful in this
  regard.
\end{itemize}

\subsection*{ACKNOWLEDGEMENTS}
M.R. is supported by a PPARC theory grant. Research conducted in
cooperation with Silicon Graphics/Cray Research utilising the Origin
3800 supercomputer (COSMOS) at DAMTP, Cambridge.  COSMOS is a UK-CCC
facility which is supported by HEFCE and PPARC.  I would like to thank
Jerry Ostriker for his valuable comments on the draft and exciting
discussions.

\appendix

\section{Generalised Profile}\label{app}

The comoving virial radius, $r_\Delta$, here is defined as the radius where
the halo mean overdensity is $\Delta$, \ie, $M_{dm} \equiv (4 \pi/3) \rho_0
\Delta r_\Delta^3$.  Here $M_{dm}$ is the mass of the halo, $\rho_0$
is the mean DM density at $z=0$ and $\Delta \approx 178$ if
$\Omega(z)=1$.  A density profile with inner slope $\alpha$, and outer
slope $\gamma$, can be written in the form
\[
\rho_{dm}(r)={\rho_0\Delta \delta_{(c,\alpha,\gamma)} \over X^\alpha(1+X)^{\gamma-\alpha}}
\]
where $X=r/r_s$, $r_s=r_\Delta/c_\Delta$ is the core
radius and
$\delta_{(c,\alpha,\gamma)}$ is a normalisation constant. The
NFW profile has $\alpha=1$ and $\gamma=3$. The mass enclosed inside the
radius, $X$, is
\begin{equation}
\begin{split}
  m_{dm}(X)=4 \pi \rho_0\Delta \delta_{(c,\alpha,\gamma)} r_s^3 \int_0^X
  {y^g \over (1+y)^{\gamma-\alpha}}{dy \over
    y} = \\
  4 \pi \rho_0\Delta \delta_{(c,\alpha,\gamma)} r_s^3 \left({X^g \over
      g}\right) {_2F_1}(g ;\gamma-\alpha; g+1; -X)
\end{split}
\label{eq:mass}
\end{equation}
where $g=(3-\alpha)$ and $_2F_1$ is the hypergeometric function. If
$\gamma=3$, the hypergeometric function simplifies as 
\[
{_2F_1}(g ;g ; g+1; -X)= g \sum_{n=0}^\infty {{\cal B}(n; g) \over
  (g+n)n} (-X)^n ,
\]
where $\cal B$ is the beta function. When $\alpha=1$ and $\gamma=3$
(NFW profile) \eq~(\ref{eq:mass}) can be integrated analytically and
the solution is
\[
m_{dm}(X)={4 \pi \over 3}\rho_0\Delta \delta_c^{NFW} r_s^3 \left[\ln(1+X)-{X \over 1+X}\right].
\]
\begin{figure}
\centerline{\psfig{figure=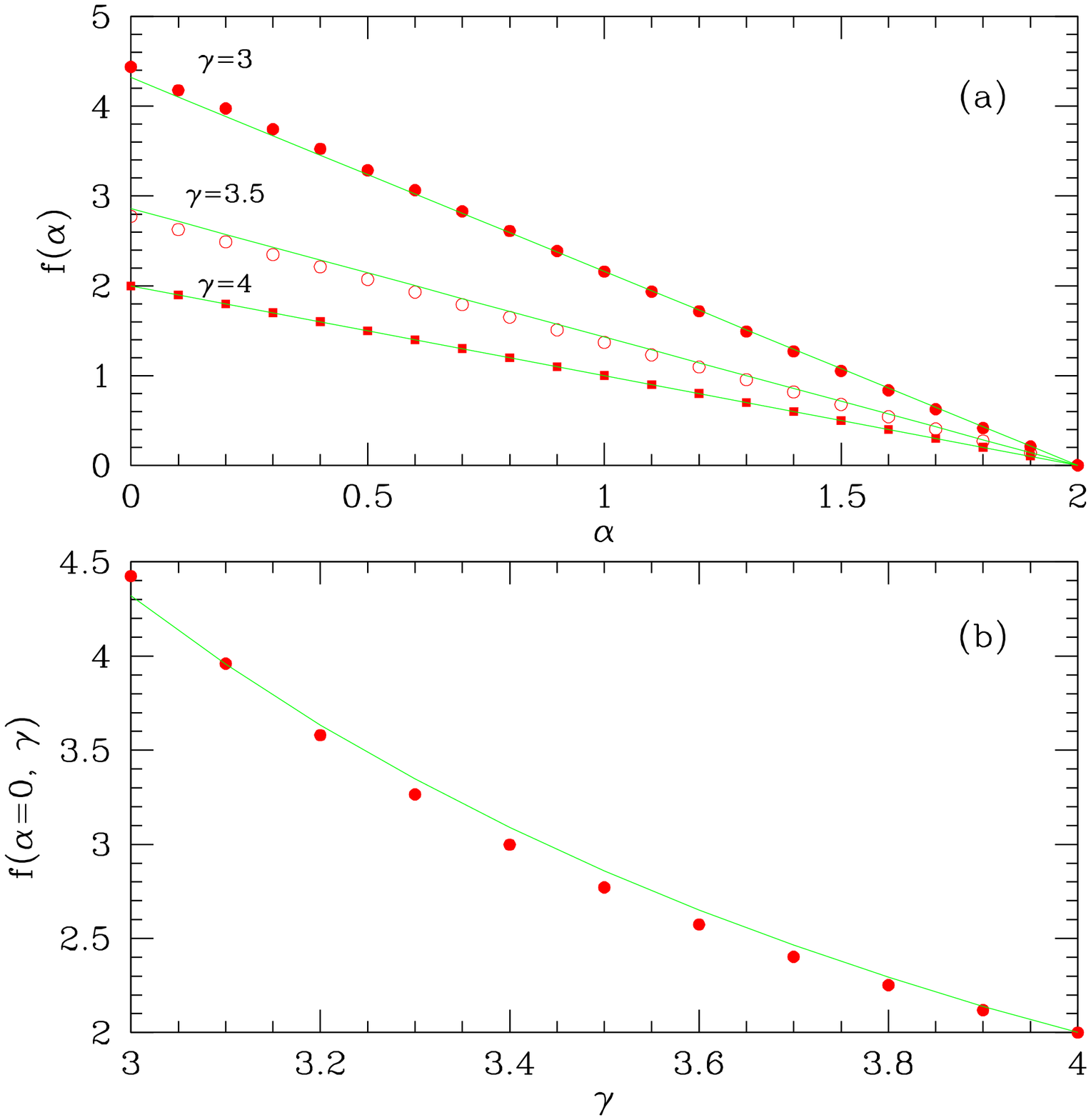,width=9cm}}
\caption{\label{fig:fit} (a) $f(\alpha,
  \gamma)=r_{max}/r_s$ as a function of $\alpha$ for $\gamma=3$
  (circles), $\gamma=3.5$ (open circles), and $\gamma=4$ (squares).
  The solid lines show the fit to the numerical solution. (b)
  $f(\alpha=0, \gamma)=r_{max}/r_s$ as a function of $\gamma$.  The
  solid lines show the fit, $f=(4/\gamma)^{2.65}(2-\alpha)$, to the
  numerical solution.}
\end{figure}
The normalisation constant $\delta_{(c, \alpha, \gamma)}$ is such
that $m_{dm}(X=c_\Delta)=M_{dm}$. For the NFW profile,
$\delta_c^{NFW}=c_\Delta^3/[\ln(1+c_\Delta)
-c_\Delta/(1+c_\Delta)]$. In the general case \eq~(\ref{eq:mass}) has
to be evaluated numerically. 
If $\gamma=3$, analytical solutions exist for $\alpha=0, 0.5, 1,
1.5$ and 2:
\[
_2F_1(-X)=
\begin{cases}
{3 \over X^3}\left[-{(1+3X/2)X \over (1+X)^2} + \ln{(1+X)}\right],~~\text{if}~\alpha=0\\
{5 \over X^{5/2}}\left[-{(1+4X/3)X^{1/2} \over (1+X)^{3/2}} + {\rm ArcSinh}(X^{1/2})\right],~~\text{if}~\alpha=0.5\\
{3 \over X^{3/2}}\left[-\left({X \over 1+X}\right)^{1/2} + {\rm ArcSinh}(X^{1/2})\right]~~\text{if}~\alpha=1.5\\
{\ln (1+X) \over X},~~\text{if}~\alpha=2
\end{cases}
\]
If $\gamma=4$ an analytical solution exists for a generic $\alpha$:
\[
_2F_1(-X)=\left({1+X \over X}\right)^\alpha[1-3(1+X)^{-1}+3(1+X)^{-2}-(1+X)^{-3}].
\]

The circular velocity is defined as
\begin{equation}
V_c(X) = \sqrt{m_{dm}(X) \over X r_s}.
\label{eq:cv}
\end{equation}
For the NFW profile \eq~(\ref{eq:cv}) has a maximum $V_c^{max} \propto
r_\Delta$ at $r_{max}=2.16r_s$. For a fixed value of $r_\Delta$,
$V^{max}_c$ depends weakly on $c_\Delta$, and $r_{max}$ is inversely
proportional to $c_\Delta$.  In the general case, $r_{max}=f(\alpha,
\gamma)r_s$ and must be calculated numerically. In \fig~\ref{fig:fit}(a)
I show the numerical solution and a simple fit to $f(\alpha,
\gamma)=r_{max}/r_s$ as a function of $\alpha$ for $\gamma=3, 3.5$ and
4.  If $\alpha=0$ the maximum, $f$, of $v_c$ can be found solving the
equation
\[
2[1-(1+f)^\gamma] +(\gamma-1)f\left[(\gamma-2)^2f^2 + \gamma f +{2\gamma \over
\gamma-1}\right]= 0.
\]
If $\gamma=4$ the solution is $f=2$, but in general the solution has
to be evaluated numerically. The simplest fit to the numerical
solution, shown in \fig~\ref{fig:fit}(b), is given by $f=(4 /
\gamma)^{2.68}$. Using this fit in conjunction with the previous for
$f(\alpha,3)$, a general fitting formula is found to be
\begin{equation}
{r_{max} \over r_s}=f(\alpha, \gamma) \approx (4 / \gamma)^{2.68}(2-\alpha),
\end{equation}
which is at least 5 percent accurate in the parameter interval $0<
\alpha <2$ and $3 < \gamma < 4$.

\bibliographystyle{/home/ricotti/Latex/TeX/mn2e}
\bibliography{/home/ricotti/Latex/TeX/archive}

\begin{thebibliography}{}

\bibitem[\protect\citeauthoryear{{Alvarez}, {Ahn} \& {Shapiro}}{{Alvarez}
  et~al.}{2003}]{AlvarezS:03}
{Alvarez} M.~A.,  {Ahn} K.,    {Shapiro} P.~R.,  2003, in M.~Reyes E. V.-S.,
  ed., "The Eighth Texas-Mexico Conference on Astrophysics" RevMexAA SC

\bibitem[\protect\citeauthoryear{{Bertschinger}}{{Bertschinger}}{1995}]{Bertsc%
hinger:95}
{Bertschinger} E.,  1995, COSMIC, GC-3 report, (astro-ph/9506070)

\bibitem[\protect\citeauthoryear{{Bertschinger} \& {Gelb}}{{Bertschinger} \&
  {Gelb}}{1991}]{Bertschinger:91}
{Bertschinger} E.,  {Gelb} J.~M.,  1991, Computers in Physics, 5, 164

\bibitem[\protect\citeauthoryear{{Bode}, {Ostriker} \& {Turok}}{{Bode}
  et~al.}{2001}]{BodeO:01}
{Bode} P.,  {Ostriker} J.~P.,    {Turok} N.,  2001, \apj, 556, 93

\bibitem[\protect\citeauthoryear{{Borriello} \& {Salucci}}{{Borriello} \&
  {Salucci}}{2001}]{BorrielloS:01}
{Borriello} A.,  {Salucci} P.,  2001, \mnras, 323, 285

\bibitem[\protect\citeauthoryear{{Chiu}, {Gnedin} \& {Ostriker}}{{Chiu}
  et~al.}{2001}]{Chiu:01}
{Chiu} W.~A.,  {Gnedin} N.~Y.,    {Ostriker} J.~P.,  2001, \apj, 563, 21

\bibitem[\protect\citeauthoryear{{Cole} \& {Lacey}}{{Cole} \&
  {Lacey}}{1996}]{Cole:96}
{Cole} S.,  {Lacey} C.,  1996, \mnras, 281, 716

\bibitem[\protect\citeauthoryear{{de Blok} \& {Bosma}}{{de Blok} \&
  {Bosma}}{2002}]{deBlok:02}
{de Blok} W.~J.~G.,  {Bosma} A.,  2002, \aap, 385, 816

\bibitem[\protect\citeauthoryear{{de Blok}, {McGaugh}, {Bosma} \& {Rubin}}{{de
  Blok} et~al.}{2001}]{deBlok:01}
{de Blok} W.~J.~G.,  {McGaugh} S.~S.,  {Bosma} A.,    {Rubin} V.~C.,  2001,
  \apjl, 552, L23

\bibitem[\protect\citeauthoryear{{Dekel}, {Devor} \& {Hetzroni}}{{Dekel}
  et~al.}{2003}]{Dekel:03}
{Dekel} A.,  {Devor} J.,    {Hetzroni} G.,  2003, MNRAS, in press
  (astro-ph/0204452)

\bibitem[\protect\citeauthoryear{{Eke}, {Navarro} \& {Frenk}}{{Eke}
  et~al.}{1998}]{Eke:98}
{Eke} V.~R.,  {Navarro} J.~F.,    {Frenk} C.~S.,  1998, \apj, 503, 569

\bibitem[\protect\citeauthoryear{{Eke}, {Navarro} \& {Steinmetz}}{{Eke}
  et~al.}{2001}]{Eke:01}
{Eke} V.~R.,  {Navarro} J.~F.,    {Steinmetz} M.,  2001, \apj, 554, 114

\bibitem[\protect\citeauthoryear{{Gnedin} \& {Bertschinger}}{{Gnedin} \&
  {Bertschinger}}{1996}]{GnedinB:96}
{Gnedin} N.~Y.,  {Bertschinger} E.,  1996, \apj, 470, 115+

\bibitem[\protect\citeauthoryear{{Huss}, {Jain} \& {Steinmetz}}{{Huss}
  et~al.}{1999}]{Huss:99}
{Huss} A.,  {Jain} B.,    {Steinmetz} M.,  1999, \apj, 517, 64

\bibitem[\protect\citeauthoryear{{Jing}}{{Jing}}{2000}]{Jing:00}
{Jing} Y.~P.,  2000, \apj, 535, 30

\bibitem[\protect\citeauthoryear{{Jing} \& {Suto}}{{Jing} \&
  {Suto}}{2000}]{JingS:00}
{Jing} Y.~P.,  {Suto} Y.,  2000, \apjl, 529, L69

\bibitem[\protect\citeauthoryear{{Klypin}, {Kravtsov}, {Valenzuela} \&
  {Prada}}{{Klypin} et~al.}{1999}]{Klypin:99}
{Klypin} A.,  {Kravtsov} A.~V.,  {Valenzuela} O.,    {Prada} F.,  1999, \apj,
  522, 82

\bibitem[\protect\citeauthoryear{{Kravtsov}, {Klypin}, {Bullock} \&
  {Primack}}{{Kravtsov} et~al.}{1998}]{Kravtsov:98}
{Kravtsov} A.~V.,  {Klypin} A.~A.,  {Bullock} J.~S.,    {Primack} J.~R.,  1998,
  \apj, 502, 48

\bibitem[\protect\citeauthoryear{{Moore}, {Governato}, {Quinn}, {Stadel} \&
  {Lake}}{{Moore} et~al.}{1998}]{Moore:98}
{Moore} B.,  {Governato} F.,  {Quinn} T.,  {Stadel} J.,    {Lake} G.,  1998,
  \apjl, 499, L5

\bibitem[\protect\citeauthoryear{{Moore}, {Quinn}, {Governato}, {Stadel} \&
  {Lake}}{{Moore} et~al.}{1999}]{Moore:99}
{Moore} B.,  {Quinn} T.,  {Governato} F.,  {Stadel} J.,    {Lake} G.,  1999,
  \mnras, 310, 1147

\bibitem[\protect\citeauthoryear{{Navarro}, {Frenk} \& {White}}{{Navarro}
  et~al.}{1996}]{Navarro:96}
{Navarro} J.~F.,  {Frenk} C.~S.,    {White} S. D.~M.,  1996, \apj, 462, 563

\bibitem[\protect\citeauthoryear{{Navarro}, {Frenk} \& {White}}{{Navarro}
  et~al.}{1997}]{Navarro:97}
{Navarro} J.~F.,  {Frenk} C.~S.,    {White} S. D.~M.,  1997, \apj, 490, 493

\bibitem[\protect\citeauthoryear{{Nipoti}, {Londrillo} \& {Ciotti}}{{Nipoti}
  et~al.}{2003}]{NipotiLC:03}
{Nipoti} C.,  {Londrillo} P.,    {Ciotti} L.,  2003, MNRAS, accepted

\bibitem[\protect\citeauthoryear{{Power}, {Navarro} \& {Steinmetz}}{{Power}
  et~al.}{2002}]{Power:02}
{Power} C.,  {Navarro} J.~F.,    {Steinmetz} M.,  2002, {The inner},
  (astro-ph/0201544) submitted

\bibitem[\protect\citeauthoryear{{Ricotti}, {Gnedin} \& {Shull}}{{Ricotti}
  et~al.}{2002a}]{RicottiGSa:02}
{Ricotti} M.,  {Gnedin} N.~Y.,    {Shull} J.~M.,  2002a, \apj, 575, 33

\bibitem[\protect\citeauthoryear{{Ricotti}, {Gnedin} \& {Shull}}{{Ricotti}
  et~al.}{2002b}]{RicottiGSb:02}
{Ricotti} M.,  {Gnedin} N.~Y.,    {Shull} J.~M.,  2002b, \apj, 575, 49

\bibitem[\protect\citeauthoryear{{Ricotti} \& {Wilkinson}}{{Ricotti} \&
  {Wilkinson}}{2003}]{RicottiW:03}
{Ricotti} M.,  {Wilkinson} M.~I.,  2003, in preparation

\bibitem[\protect\citeauthoryear{{Salucci} \& {Burkert}}{{Salucci} \&
  {Burkert}}{2000}]{SalucciB:00}
{Salucci} P.,  {Burkert} A.,  2000, \apjl, 537, L9

\bibitem[\protect\citeauthoryear{{Spergel} \& {Steinhardt}}{{Spergel} \&
  {Steinhardt}}{2000}]{Spergel:00}
{Spergel} D.~N.,  {Steinhardt} P.~J.,  2000, Physical Review Letters, 84, 3760

\bibitem[\protect\citeauthoryear{{Subramanian}, {Cen} \&
  {Ostriker}}{{Subramanian} et~al.}{2000}]{SubramanianCO:00}
{Subramanian} K.,  {Cen} R.,    {Ostriker} J.~P.,  2000, \apj, 538, 528

\bibitem[\protect\citeauthoryear{{Syer} \& {White}}{{Syer} \&
  {White}}{1998}]{SyerW:98}
{Syer} D.,  {White} S.~D.~M.,  1998, \mnras, 293, 337

\bibitem[\protect\citeauthoryear{{van den Bosch}, {Robertson}, {Dalcanton} \&
  {de Blok}}{{van den Bosch} et~al.}{2000}]{vandenBosch:00}
{van den Bosch} F.~C.,  {Robertson} B.~E.,  {Dalcanton} J.~J.,    {de Blok}
  W.~J.~G.,  2000, \aj, 119, 1579

\bibitem[\protect\citeauthoryear{{Zhao}}{{Zhao}}{1996}]{Zhao:96}
{Zhao} H.,  1996, \mnras, 278, 488

\end{thebibliography}

\label{lastpage}
\end{document}